# A deep survey of the AKARI North Ecliptic Pole Field
## I. *WSRT* 20 cm Radio survey description, observations and data reduction

Glenn J. White[1,2], Chris Pearson[2,3,1], Robert Braun[4], Stephen Serjeant[1], Hideo Matsuhara[5], Toshinobu Takagi[5], Takao Nakagawa[5], Russell Shipman[6], Peter Barthel[7], Narae Hwang[8], Hyung Mok Lee[9], Myung Gyoon Lee[9], Myungshin Im[9], Takehiko Wada[5], Shinki Oyabu[5], Soojong Pak[9], Moo-Young Chun[9], Hitoshi Hanami[10], Tomotsugu Goto[11,8], Seb Oliver[12]

[1] Department of Physics and Astronomy, The Open University, Walton Hall, Milton Keynes, MK7 6AA, UK
[2] Space Science and Technology Department, STFC Rutherford Appleton Laboratory, Chilton, Didcot, Oxfordshire, OX11 0QX, UK
[3] Institute for Space Imaging Science, University of Lethbridge, Lethbridge, Alberta, T1K 3M4, Canada
[4] Australia Telescope National Facility, CSIRO, Marsfield NSW 2122, Australia
[5] Institute of Space and Astronautical Science, JAXA, Yoshino-dai 3-1-1, Sagamihara, Kanagawa 229-8510, Japan
[6] SRON Netherlands Institute for Space Research, PO Box 800, 9700 AV Groningen, The Netherlands
[7] Kapteyn Astronomical Institute, University of Groningen, P.O. Box 800, 9700 AV Groningen, Netherlands
[8] National Astronomical Observatory of Japan, Osawa, Mitaka, Tokyo 181-8588, Japan
[9] Astronomy Program, Department of Physics and Astronomy, Seoul National University, Seoul 151-747, Korea
[10] Physics Section, Faculty of Humanities and Social Sciences, Iwate University, Morioka 020-8550, Japan
[11] Institute for Astronomy, University of Hawaii, 2680 Woodlawn Drive, Honolulu, HI 96822, USA
[12] Department of Physics & Astronomy, School of Science and Technology, University of Sussex, Falmer, Brighton BN1 9QH, UK

**ABSTRACT**

*Aims.* The Westerbork Radio Synthesis Telescope, *WSRT*, has been used to make a deep radio survey of an $\sim 1.7$ degree$^2$ field coinciding with the AKARI North Ecliptic Pole Deep Field. The observations, data reduction and source count analysis are presented, along with a description of the overall scientific objectives.
*Methods.* The survey consisted of 10 pointings, mosaiced with enough overlap to maintain a similar sensitivity across the central region that reached as low as 21 $\mu$Jy beam$^{-1}$ at 1.4 GHz.
*Results.* A catalogue containing 462 sources detected with a resolution of $17.0'' \times 15.5''$ is presented. The differential source counts calculated from the *WSRT* data have been compared with those from the shallow $VLA - NEP$ survey of Kollgaard et al. 1994, and show a pronounced excess for sources fainter than $\sim 1$ mJy, consistent with the presence of a population of star forming galaxies at sub-mJy flux levels.
*Conclusions.* The AKARI North Ecliptic Pole Deep field is the focus of a major observing campaign conducted across the entire spectral region. The combination of these data sets, along with the deep nature of the radio observations will allow unique studies of a large range of topics including the redshift evolution of the luminosity function of radio sources, the clustering environment of radio galaxies, the nature of obscured radio-loud active galactic nuclei, and the radio/far-infrared correlation for distant galaxies. This catalogue provides the basic data set for a future series of paper dealing with source identifications, morphologies, and the associated properties of the identified radio sources.

## 1. Introduction

Deep radio and far-infrared (far-IR) surveys are useful to study the global properties of extragalactic source populations in the early Universe; to measure the evolution of AGN's and starburst galaxies at early epochs; and to understand the cosmic history of star formation. Recently, the Japanese AKARI infrared satellite has made deep surveys close to the North and the South Ecliptic Poles. These regions have relatively low line of sight extinction (to the distant Universe) and low hydrogen column densities, which are important if objects at large distances are to be detectable at optical and infrared wavelengths. To support the AKARI North Ecliptic Pole (*NEP*) Survey (Matsuhara et al. 2006, Wada et al. 2008), this region has been observed using the Westerbork Radio Synthesis Telescope (*WSRT*).

The observational results of the *WSRT* − *AKARI* − *NEP* survey will be presented in three papers: a) the present paper presents the basic radio survey, source catalogues, radio source counts and statistics; b) a second paper will report the results from cross-correlation between the *WSRT* radio observations and the infrared source catalogue from the AKARI North Ecliptic Pole survey; and c) the third paper will present optical identifications from a cross-correlation between the *WSRT* radio survey and deep optical imaging made

*Send offprint requests to*: g.j.white@open.ac.uk

using the Canada France Hawaii 3.6 metre (CFHT) and SUBARU 8 metre telescopes, and will address the more global objectives of the survey stated above.

## 2. Multi-wavelength observations

The two Ecliptic Poles are amongst the deepest exposure regions that have been observed by many infrared satellite missions, and provide a wealth of data about the distant source populations, for example the surveys of IRAS (Hacking, Condon & Houck 1987, Aussel et al. 2000), ISO (Stickel et al. 1998, Aussel et al. 2000), COBE (Bennett et al. 1996), and ROSAT (Mullis et al. 2001, 2003). Other surveys of this region at radio wavelengths have been made with the *VLA* (Kollgaard et al. 1994, Brinkmann et al. 1999 at 20 and 91 cm); Westerbork: Rengelink et al. (1997); Effelsberg 100 metre telescope (Loiseau et al. 1988); and in 2.7 GHz surveys by Condon & Broderick (1985, 1986) and Loiseau et al. (1988); at optical/IR wavelengths (Gaidos, Magnier & Schechter 1993 and Kümmel et al. 2000, 2001); and at X-ray wavelengths using ROSAT by Henry et al. (2001) and Mullis et al. (2001). The area around the *NEP* has a moderate/low level of HI emission $\sim 4.3 \times 10^{20}$ cm$^{-2}$ (Elvis, Lockman & Fassnacht 1994). This corresponds to a line of sight extinction $A_v \sim 0.2 - 0.5$ magnitudes, favouring very deep optical and near-infrared observations because of the low level of foreground extinction (Zickgraf et al. 1997). Optical and infrared surveys provide key information to help to understand the source populations of the *NEP* region, in particular the AKARI mission and its supporting ancillary programmes have included two deep 2.4-24 $\mu$m wavelength surveys at the North Ecliptic Pole (*NEP*): a) covering a 0.4 deg$^2$ circular area (known as *NEP*-Deep - see Matsuhara et al. 2006), and b) a wide and shallow 2.4-24 $\mu$m survey covering a 5.8 deg$^2$ circular area surrounding the *NEP*-Deep field (also known as *NEP*-Wide – Lee et al. 2009).

Optical, radio, X-ray and infrared surveys provide essential support to the interpretation of deep extragalactic radio surveys. A shallow *VLA* 20 cm survey of the *NEP* region was made by Kollgaard et al. (1994), which covered an area of 29.3 deg$^2$. The Kollgaard survey reported 2435 radio sources with flux densities ranging from 0.3–1000 mJy, observed with a 20″ beam and 1$\sigma$ noise $\sim 60$ $\mu$Jy per beam at the centre of the survey field. A comparison between this radio survey and the NASA Extragalactic database, and with other catalogues (including the ROSAT X-ray catalogue), resulted in the identification of $\sim 20\%$ of the sources, with $\sim 6\%$ of the sources found to be extended with diameters $\geq 30''$. A 2.7 GHz survey of the region was made by Condon & Broderick (1985, 1986). Between 1 and 150 mJy, the slope of the log $N$ – log $S$ relationship was 0.68±0.03. An even larger area of 570 degrees$^2$ was observed at 325 MHz using the *WSRT* telescope by Rengelink et al. (1997) in the *WENSS* survey (beam size 54″), which resulted in the detection of more than 11,000 sources. The source populations have already in this region include galaxy clusters (Gioia et al. 2003, 2004, Hwang et al. 2007, Goto et al. 2008), radio galaxy clusters (Branchesi et al. 2006), stars (Pretorious et al. 2007, Micela et al. 2007, Affer et al. 2008), X-ray sources (Voges et al. 2001, Henry et al. 2001, 2006) and infrared sources (Kümmel et al. 2000).

## 3. *WSRT* Observations

The radio observations presented in this paper were observed during 2004 with the *WSRT* operated at 20 cm wavelength. The array included fourteen 25 m telescopes arranged in a 2.7 km east-west configuration, with signals processed using a digital continuum 2-bit back end consisting of eight 20 MHz bandwidth sub-bands across the frequency range 1301–1461 MHz. The *NEP* observations were interleaved with observations of the intensity, polarisation and phase calibration sources 3C147 and 3C286, which past experience suggested should lead to a flux density calibration accuracy of better than 5%. The survey mosaic was made from 10 discrete pointings that were positioned on an hexagonal grid, with beam spacings at the 70% point of the 36.2′ primary beam full width half maximum (FWHM) diameter, each observed as a full 12 hour track. This observing strategy was adopted to provide a relatively uniform noise background of less than ±10% over the most sensitive part of the surveyed area (see Prandoni et al. 2006) for a full treatment of mosaicing strategies, who show that this results in mosaiced noise variations of $\leq 5\%$), where a 1 $\sigma$ source detection sensitivity of point sources as low as 21 $\mu$Jy per beam was achieved. The J2000 coordinate system is used throughout this paper. Experience at the WSRT suggests that the interpolation and coplanarity (also known as faceting) processing in the mosaicing step should not introduce errors in excess of 0.1 arc seconds for a relatively small mosaic of this size.

The observations were reduced and calibrated using standard tasks in the *AIPS* software package. The data sets were uniformly of high quality, with only a few percent of the visibilities having to be flagged out, mostly due to low level radio interference. Each pointing was mapped onto a regular grid with 4″ pixels using a multi-frequency synthesis approach to minimise bandwidth smearing. Adjacent pointings were co-added to the FWHM point (Condon et al. 1998, Huynh et al. 2005). After a first iteration, model components with a flux density of more than $\sim 1$ mJy beam$^{-1}$ were used for phase and amplitude self-calibration, to correct for residual phase and amplitude errors. The data were then re-imaged and cleaned with $\sim 2000$ clean components, at which point the side lobes of most of the strong sources were found to be below the noise level. There was however a particular problem toward the central position in the mosaic where the prominent galactic planetary nebula, the Cat's Eye Nebula (NGC 6543) lies close to the field centre. Since this is slightly extended at radio wavelengths, it presented a particular challenge to the data reduction and cleaning, and ultimately limited the *rms* noise level achievable in the immediate vicinity to be several times the thermal limit. However, the number of pixels affected was very small ($\leq 0.1\%$ of the total), and a correction for this was made during the differential source count calculation presented in Section 7.

After reduction of the individual pointings, the maps were individually intensity corrected using a model of the primary beam, and then mosaiced together into a final image using the *AIPS* task *LTESS*, to make a linearly combined mosaic, correcting for the individual primary beam patterns, and optimizing the signal to noise ratio. The pixels at the edges of the mosaiced region have higher noise uncertainties compared to those at the centre of the merged field because of the primary beam profile, and the mosaic-ing strategy. The mosaiced, primary beam corrected image of the high sensitivity region of the *WSRT* image is shown in Fig 1.

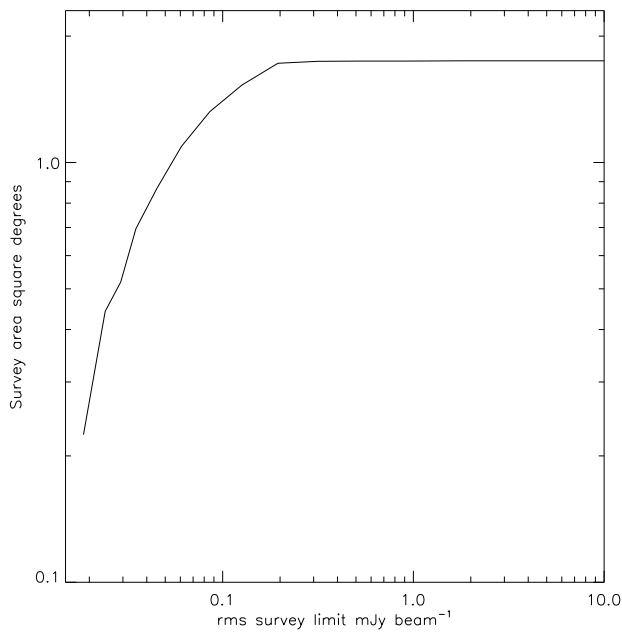

**Fig. 2.** The horizontal axis shows the typical 1 $\sigma$ noise level radially averaged as a function of the (vertical axis) total areal coverage.

Several automated source extraction and cataloging routines were tested, including the *AIPS* task *SAD* and the *MIRIAD* tasks *IMSAD* and *SFIND* (Sault, Teuben & Wright 1995), but the latter task was eventually adopted as the extraction task of choice for reasons that will be discussed in Section 4. A quantitative comparison between *SFIND*, *SAD*, *IMSAD*, and *SEXTRACTOR* has already been presented by Hopkins et al. (2002) to which the reader is referred for a rigorous treatment of noise and error estimates relevant to this paper.

The final restored beam size in the mosaic after all of the associated processing steps was $17.0'' \times 15.5''$ at position angle 0 degrees. The most sensitive part of the survey field had a 1 $\sigma$ *rms* sensitivity of 21 $\mu$Jy beam$^{-1}$ in the centre of the map, increasing to $\geq$100 $\mu$Jy beam$^{-1}$ toward the edges of the field, because of the primary beam attenuation correction. It was therefore not possible to use the same detection threshold across the whole of the mosaiced region. Furthermore, flux densities measured toward the image edges were increasingly affected by uncertainties in the primary beam model, and consequently the image analysis was restricted to those sources which lie in regions where the theoretical sensitivity is below 60 $\mu$Jy beam$^{-1}$ for noise considerations, and to mitigate other biases such as bandwidth smearing so as not to affect the source intensities by more than a few percent. To measure the noise, estimates of the rms errors were estimated using *SEXTRACTOR*, and separately using the *MIRIAD* task *SFIND*. The detection sensitivity is shown in Figure 2, with similar results being obtained in *SEXTRACTOR* and in *SFIND*.

## 4. Source Catalogue

The *NEP* mosaic has a non-uniform and continuously varying noise level, a complex mosaicing strategy, and locally elevated noise levels around the few bright sources such as the Cat's Eye Nebula, and it is clear that source detection using an uniform flux threshold over the whole primary beam corrected image is not the optimal approach. Source detection in this case is better determined using locally determined noise levels - an approach that has already been used in other studies to improve the efficacy of their source detection catalogues (e.g. Morganti et al. 2004).

The source catalogue in this paper was built using the *MIRIAD* task *SFIND*. This is a method for identifying source pixels, where the detected sources are drawn from a distribution of pixels with a robustly known chance of being falsely drawn from the background (see Hopkins et al. 1999, 2002 and Morganti et al. 2004) for a complete description justifying the adoption of this technique. *SFIND* robustly characterises the fraction of expected pixels more rigorously than from a traditional sigma-clipping criterion - which is known to suffer limitations at lower signal-to-noise levels. Noise estimation is implemented in the image plane by dividing the image into small square regions within which the mean and *rms* noise level are estimated by fitting a Gaussian to the pixel histogram in each region. The image is then normalised by subtracting the mean and dividing by the *rms* within each region, resulting in an image where pixel values are specified in units of the local *rms* noise level $\sigma$. *SFIND* uses a statistical technique, the false discovery rate (FDR), which assigns a threshold based on an acceptable rate of false detections (Hopkins et al. 2002). We followed Hopkins et al. (2002) by adopting an FDR value of 2%. Each of the sources identified by *SFIND* were visually inspected to remove any obvious mis-identifications. Comparison with independent catalogues derived using the *MIRIAD* task *IMSAD* (with a 7 $\sigma$ clip), and with one derived using *SEXTRACTOR* with a locally defined background *rms* were almost identical with the *SFIND* catalogue.

A sample from the final source catalogue is presented in Table 1, and completely in the electronic on-line version of this paper.

The positional accuracy listed in the Table 1 is relative to the self-calibrated and bootstrapped reference frame described in Section 3, after mitigating the various effects mentioned above. Several other effects that can affect the positions include the mosaicing process (which we have discussed in Section 3; the signal-to-noise of the detected sources (presented in Table 1; and other observational effects that bias the positions or sizes of sources, which are discussed in Section 5. We also present an estimate of source dimensions estimated by deconvolving the measured sizes from the synthesized beam, reporting those more than double the synthesised beam size. Although is possible to model the source sizes in a more exact way (for example following the approach of Oosterbaan (1978), we only use the present source size data as a guide to whether the sources are either extended, or likely to be multi-component sources. A more detailed discussion of the WSRT source sizes will be made in a future paper that combines the present data set with higher resolution observations of the *NEP* field with the GMRT Telescope (Sedgwick et al. 2010). To test the accuracy of the radio reference frame, WSRT sources with a peak signal to noise ratio $\geq 10$ which could be identified with compact optical galaxies from a deep SUBARU image (referenced from HST Guide Star positions) were found to have positional offsets within $\sim 2''$ of each other, randomly distributed around the nominal radio position. Further quantitative discussion of the optical identifications, and the radio-optical frame registration determined from a larger selection of optically identified sources will be presented in Paper 2, which is dedicated to the optical/infrared identifications from this survey (White et al. *in preparation*).

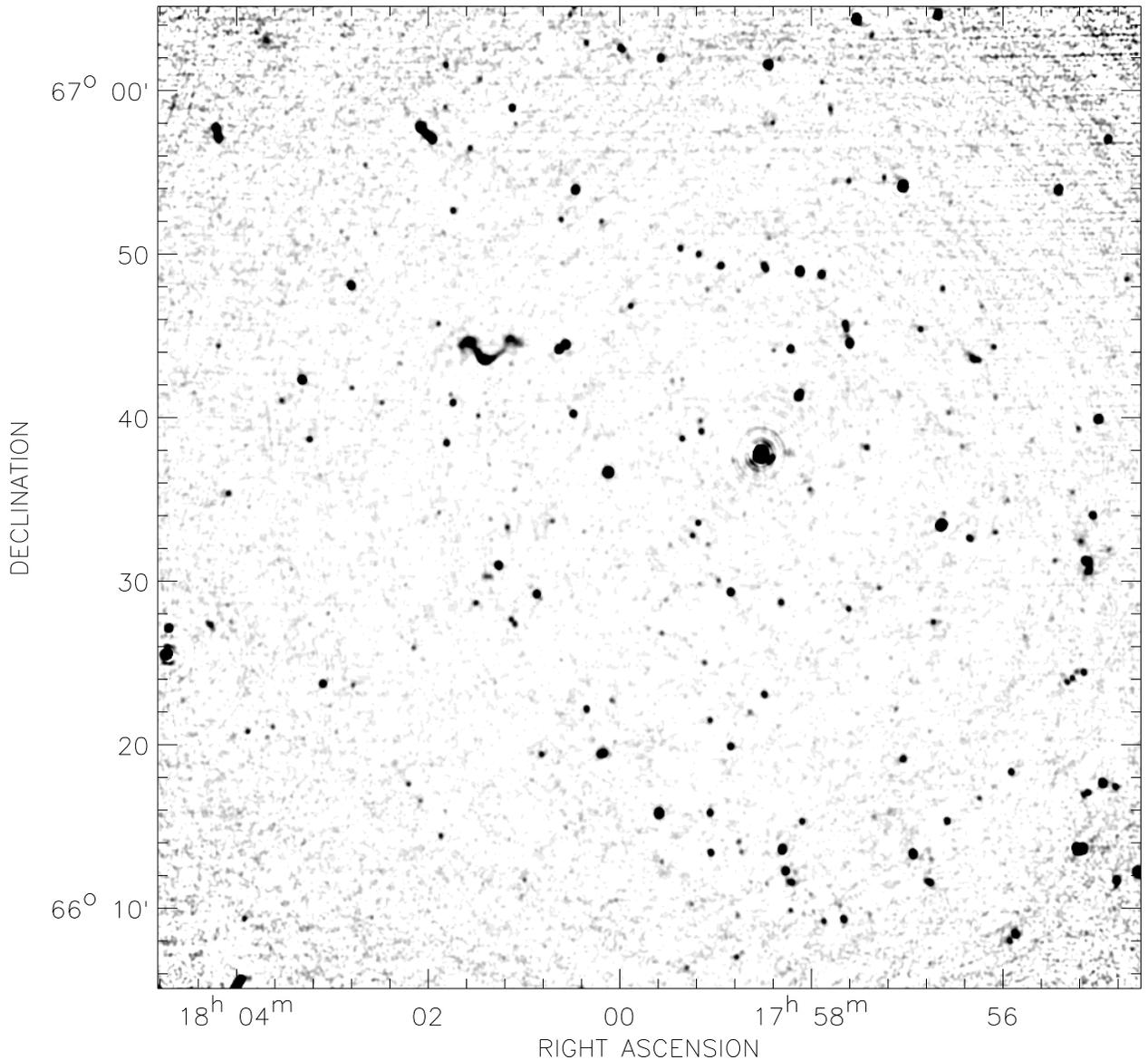

**Fig. 1.** The central 1 square degree area of the *WSRT* 20 cm map, corrected for the primary beam of the antenna. The bright Cats Eye Nebula lies $\sim 5'$ to the right of the centre of the map, where some faint residual sidelobe structure can be seen as part of a circular arc. The local noise levels are slightly elevated close to this source.

## 5. Flux Accuracy and Error Estimates

The observations from a radio synthesis array must be corrected for various instrumental effects: *a*) the primary beam response of the antenna elements; *b*) time-average smearing due to the finite integration time; *c*) chromatic aberration resulting from the finite bandwidth (Bridle & Schwab 1989; Cotton 1989), and *d*) incompleteness at low signal to noise levels. We briefly describe the approach we have taken, below.

**Table 1.** The source catalogue (the full version is available as Supplementary Material in the on line version of this article). The source parameters listed in the catalogue are: (1) a short form running number, (2) the source name, referred to in this paper as $NEP$ followed by the RA/Dec encoding (e.g. NEP175431+663536), (3,4) the source Right Ascension and Declination (J2000) referenced from the self-calibrated reference frame, (5,6) the RA and Dec errors in arc seconds, (7,8) the peak flux density, $S_{peak}$, and its associated rms error, (9,10) the integrated flux densities, $S_{total}$ and their associated errors, (11, 12, 13) the major and minor axes of the fitted Gaussian source profile and orientation (major and minor axis full width at half maximum, and position angle measured east of north.

| Running number | Source name | RA hh:mm:ss.s | DEC dd:mm:ss.s | δRA ″ | δDEC ″ | $S_{peak}$ mJy beam$^{-1}$ | $S_{peak}$ error mJy beam$^{-1}$ | $S_{total}$ mJy | $S_{total}$ error mJy | $θ_{maj}$ ″ | $θ_{min}$ ″ | PA ° |
|---|---|---|---|---|---|---|---|---|---|---|---|---|
| (1) | (2) | (3) | (4) | (5) | (6) | (7) | (8) | (9) | (10) | (11) | (12) | (13) |
| 1 | NEP175121+663645 | 17:51:21.7 | +66:36:45.1 | 1.39 | 0.22 | 1.219 | 0.270 | 1.932 | 0.295 | 28.0 | | -81.6 |
| 2 | NEP175140+665038 | 17:51:41.0 | +66:50:38.1 | 0.01 | 0.01 | 2.464 | 0.364 | 1.461 | 0.367 | | | |
| 3 | NEP175147+663124 | 17:51:47.8 | +66:31:24.1 | 0.05 | 0.04 | 1.138 | 0.177 | 0.829 | 0.182 | | | |
| 4 | NEP175214+665054 | 17:52:14.7 | +66:50:54.2 | 0.05 | 0.11 | 2.615 | 0.257 | 5.297 | 0.288 | 24.2 | 13.0 | -0.7 |
| 5 | NEP175231+662738 | 17:52:31.9 | +66:27:38.3 | 0.38 | 0.28 | 0.633 | 0.148 | 1.181 | 0.158 | 20.6 | 14.8 | -54.6 |
| 6 | NEP175248+662713 | 17:52:48.6 | +66:27:13.9 | 0.01 | 0.01 | 0.886 | 0.102 | 0.885 | 0.103 | | | |
| 7 | NEP175254+663144 | 17:52:54.5 | +66:31:44.1 | 0.14 | 0.08 | 2.774 | 0.131 | 7.340 | 0.184 | 29.7 | 17.4 | -86.5 |
| 8 | NEP175256+663148 | 17:52:56.4 | +66:31:48.4 | 0.74 | 0.10 | 2.078 | 0.131 | 7.709 | 0.204 | 59.6 | 9.7 | 85.5 |
| 9 | NEP175305+663929 | 17:53:06.0 | +66:39:29.8 | 0.02 | 0.01 | 1.708 | 0.121 | 2.216 | 0.131 | | | |
| 10 | NEP175307+663213 | 17:53:07.6 | +66:32:13.2 | 0.02 | 0.02 | 3.735 | 0.131 | 8.035 | 0.202 | 21.1 | 17.4 | 8.4 |
| 11 | NEP175313+661949 | 17:53:13.9 | +66:19:49.7 | 0.00 | 0.00 | 6.197 | 0.148 | 8.574 | 0.186 | | | |
| 12 | NEP175321+661249 | 17:53:21.5 | +66:12:49.1 | 0.04 | 0.03 | 1.403 | 0.149 | 1.673 | 0.153 | | | |
| 13 | NEP175330+662831 | 17:53:30.0 | +66:28:31.5 | 0.00 | 0.00 | 1.056 | 0.086 | 1.248 | 0.090 | | | |
| 14 | NEP175331+662726 | 17:53:31.3 | +66:27:26.6 | 0.00 | 0.00 | 26.929 | 0.086 | 32.232 | 0.147 | | | |

### 5.1. Time-average smearing

The data were observed using integration times of 60 seconds, which was estimated to lead to a reduction in the flux of point sources of no more than $\sim 1\%$ for a point source $10'$ from the field centre, and it is believed that this does not play a dominant effect in determining source sizes.

### 5.2. Chromatic aberration

To correct for bandwidth smearing, the radio analog of optical chromatic aberration, we inserted 500 artificial point source models into the $uv$-data with peak values from 5–50 $\sigma$ using the $AIPS$ task $UVCON$. This data were processed in a similar way to the $NEP$ field, and $SFIND$ was used to recover the source intensities and measure the noise uncertainties. There was no evidence significant variation of the source intensities with position in the mosaic, which is similar to the conclusion of Prandoni et al. (2000b) for a similar set of $ATCA$ data.

### 5.3. Clean bias

Radio surveys can be affected by a 'clean bias' effect, where a systematic under estimation of the peak and total source fluxes (Becker et al. 1995, White et al. 1997; Condon et al. 1998) is a consequence of redistribution of the flux from point sources to noise peaks in the image. Prandoni et al. (2000a,b) show that it is possible to mitigate this bias if the $CLEAN$ing process is stopped well before the maximum residual flux has reached the theoretical noise level. Following Garrett et al. (2000), we set the cleaning limit at 5 times the theoretical noise to mitigate against this effect.

### 5.4. Resolution bias

Resolution bias is an effect in which the peak flux densities of weak extended sources fall below the chosen detection limit, yet still have total integrated flux densities that are above the survey limit. In other studies, a 3% correction was required for source counts below 1 mJy (Moss et al. 2007, Garn et al. 2008), although no resolution correction was applied to brighter sources. This effect reduces the number of faint sources in differential source counts (see for example Hopkins et al. 2002), and we do not consider that it has a significant effect on our data reduction methodology.

### 5.5. Eddington bias

Since the source counts rise strongly with decreasing flux density, more sources will have their true fluxes 'boosted' by the effect of noise, than those that are 'reduced' at higher flux densities (see discussion in Coppin et al. 2006). To examine the effect of this, a population of 'test' point sources were added into a single field, uncorrected for the primary beam response using the $AIPS$ task '$UVMOD$', and processed and extracted in the same way as the unmosaiced survey data, with the difference between the detected counts, and those inputs, providing an estimate of the net amount of up-scattering. The effect of this was only significant in the lowest flux bin, and led to an overestimate of the source count by 16% (the boundaries of the lowest flux bin were $\sim 5\%$ above the formal survey limit. This value should be compared with the value estimated by Moss et al. (2007) of 21%, which goes slightly closer to their formal survey limit. Consequently in later analysis the counts in the lowest flux bin (110–125 $\mu$Jy were 'deboosted' by 16%. This correction has a negligible effect for fluxes above this limit, and it is safe to ignore it.

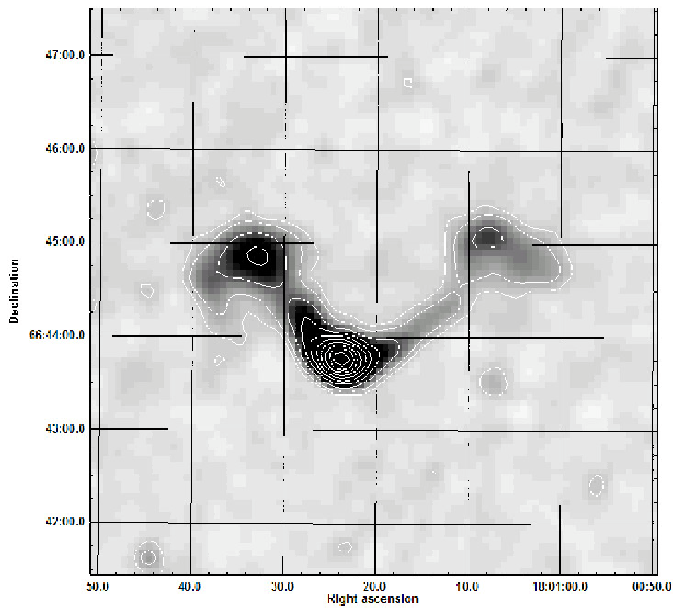

**Fig. 3.** The extended radio source centred on the $S_{peak}$ =12.3 mJy source NEP180123+664346. The contours are spaced at square root intervals for display purposes with the lowest level at 0.00010, and subsequent levels at 0.00023, 0.00063, 0.00129, 0.00222, 0.00342, 0.00488, 0.00661, 0.00861, 0.01086, 0.0122 Jy beam$^{-1}$. The two jets are strongly bent, suggesting that the source is moving through a relatively dense medium, such as that associated with a galaxy cluster.

### 5.6. Component extraction

In the terminology of this paper a radio component is described as a region of radio emission represented by a Gaussian shaped object in the map. Close radio doubles are represented by two Gaussians and are deemed to consist of two components, which make up a single source. A clear case of a very extended radio source is shown in Figure 3, and a selection of other sources with multiple components is shown in Figure 4.

### 5.7. Resolved sources

Although it may seem relatively straightforward to calculate the density of sources as a function of the flux density, the distribution of angular sizes as a function of the flux density may also bias the results. It was assumed that the median sizes below 1 mJy remain approximately constant as a function of the flux density with those at higher flux levels. Fomalont et al. (2006) find that 8±4% of the $\mu$Jy sources have sizes greater than 4″. For low signal-to-noise ratio detections, Gaussian fitting routines may be significantly affected by noise spikes, leading to errors in the estimated widths and flux densities of the sources (Moss et al. 2007)). This is one of the reasons for adopting the *SFIND* source extraction methodology in this paper. The ratio $S_{total}/S_{peak} = (\theta_{min}\,\theta_{maj})/(b_{min}\,b_{maj})$ where $\theta_{min}$ and $\theta_{maj}$ are the major and minor axes of the detected source and b$_{min}$ and b$_{maj}$ are the major and minor axes of the restoring beam. The flux density ratio may be used to discriminate between unresolved sources and those which are much larger than the beam (see Prandoni et al. 2006). In Figure 5, the ratio of the flux densities to the signal-to-noise ratio ($S_{peak}/\sigma_{local}$) is plotted for all sources above a 6 $\sigma$ local threshold. The biases introduced by using different thresholds have been modelled by Prandoni et al. (2000a,b), Owen et al. (2008) and Fomalont et al. (2006), which suggest that the biases are most prevalent below an ∼ 5–6 $\sigma$ (sigma-clip) threshold. To identify sources for which $S_{total}/S_{peak}$ <1, a functional form of the curve f(x) = 1.0 ± 3.22/x was plotted on Fig 5 to define the point where 90% of the ≥6$\sigma$ detections with $S_{total}/S_{peak}$ <1 lie above the curve (this is similar to the ratio adopted by Prandoni et al. 2000a, b). Reflecting this curve about $S_{total}/S_{peak}$ = 1 shows those sources which lie between the two curves, and which are considered to be unresolved.

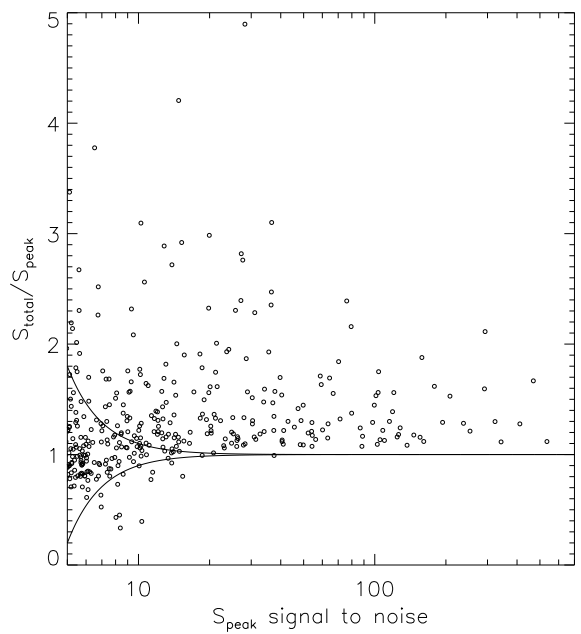

**Fig. 5.** Ratio of the integrated flux $S_{total}$ to the peak one $S_{peak}$ as a function of the source signal-to-noise. The lower and upper envelopes (dashed lines) of the flux ratio distribution are shown, along with small dots showing the unresolved sources, and larger filled circles indicating extended sources. It is likely that one of the measurements of $S_{total}$ or $S_{peak}$ where ($S_{total}$ / $S_{peak}$) ¡ 1, have been affected by noise, to the extent that the value of the ration is below unity - which mainly happens for the weaker sources.

In Figure 5 the flux ratio is shown as a function of the signal-to-noise for all the sources (or source components) in the *NEP* catalogue.

The flux density ratio shows a skewed distribution, where the tail toward high flux ratios is due to the presence of extended sources. Values for $S_{total}/S_{peak}$ < 1 result from the effect of noise in affecting the source sizes (see Section 4). To establish a criterion for extension, such noise errors have to be taken into account. The lower envelope of the flux ratio distribution (the curve containing 90% of the sources) was determined, and mirrored it on its side (upper envelope in Figure 5), so that unresolved sources should lie below the upper envelope. The upper envelope can be characterised by the equation from Huynh et al. (2005) that was found to characterise the 90% envelope of sources where $S_{\rm total} \leq S_{\rm peak}$, and a 5 $\sigma$ cut off to the peak fluxes was adopted:

$$\frac{S_{\rm total}}{S_{\rm peak}} = 1 + \left[\frac{10}{(S_{\rm peak}/\sigma_{fit})^3}\right] \quad (1)$$

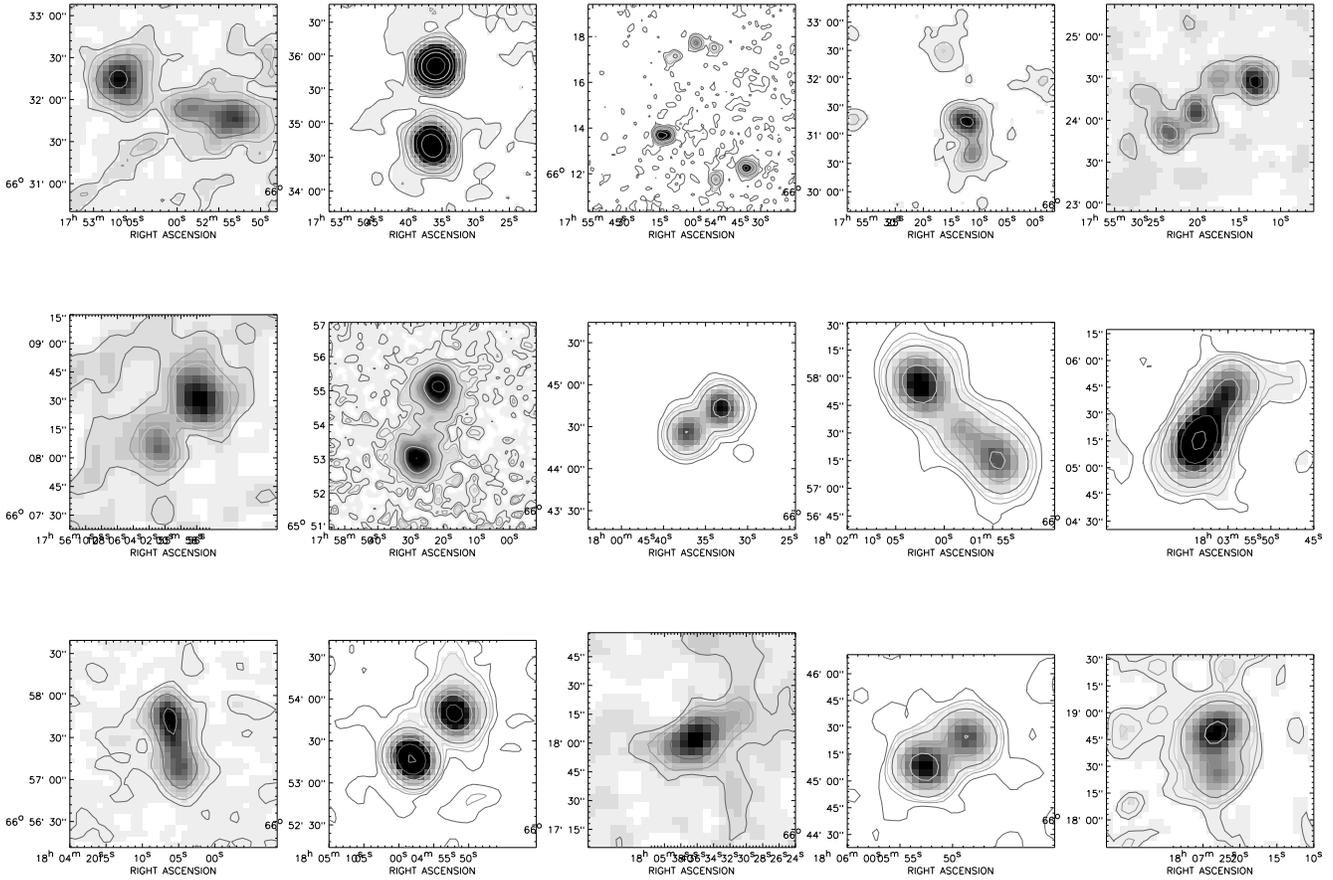

**Fig. 4.** Regions showing complex or extended structure. The vertical scale is Declination. The contours are at 0.0001, 0.0003, 0.0005, 0.001, 0.003, 0.006, 0.012, 0.024, 0.048 and 0.096 Jy beam$^{-1}$ respectively.

It is worth noting that the envelope does not converge to unity at large signal-to-noise values. This is due to the radial smearing effect which systematically reduces the *peak* fluxes, leading to larger $S_{\rm total}/S_{peak}$ ratios. From Figure 5 we estimate the smearing effect to be 8% on average for the WSRT data (a similar effect has been reported by Prandoni et al. (2000a, b) comparing *ATCA* and *VLA* data in an overlapping field, where they report applying a 5% correction to their data). The fluxes in the Table have been corrected for these effects as described in Section 5.11.

Radio sources are often made up of multiple components, as seen earlier in Figure 4. The source counts need to be corrected for this, so that the fluxes of physically related components are summed together, rather than being treated as separate sources. Magliocchetti et al. (1998) have proposed criteria to identify the double and compact source populations, by plotting the separation of the nearest neighbour of a source against the summed flux of the two sources, and selecting for objects where the ratio of their fluxes, $f_1$ and $f_2$ is in the range $0.25 \leq f_1/f_2 \leq 4$. In Figure 6 the sum of the fluxes of nearest neighbours are plotted against their separation.

The dashed line marks the boundary satisfying the separation criterion defined by Huynh et al. (2005):

$$\theta = 100 \left[\frac{S_{\rm total}({\rm mJy})}{10}\right]^{0.5} \quad (2)$$

where $\theta$ is in arc seconds. The 82 radio components in the present survey (i.e. 18% of the 462 catalogued entries) should be considered to be a part of double or multiple sources, and this will be taken account of in the differential source counts discussed later. A further correction for the incompleteness due to extended sources (Windhorst et al. 1993, Bondi et al. 2003) was considered, but found to have a negligible effect on the final catalogue, because of the relatively large beam in the present survey.

### 5.8. Positional accuracy

Noise fluctuations limit the *rms* positional uncertainty in each of the fitted sky coordinates ($\Delta$RA or $\Delta$Dec) of a faint point source with an rms brightness fluctuation $\sigma$ and FWHM resolution $\theta$ to (following Rengelink et al. 1997):

$$\sigma_p \approx \frac{\sigma\theta}{2S_{peak}} \qquad (3)$$

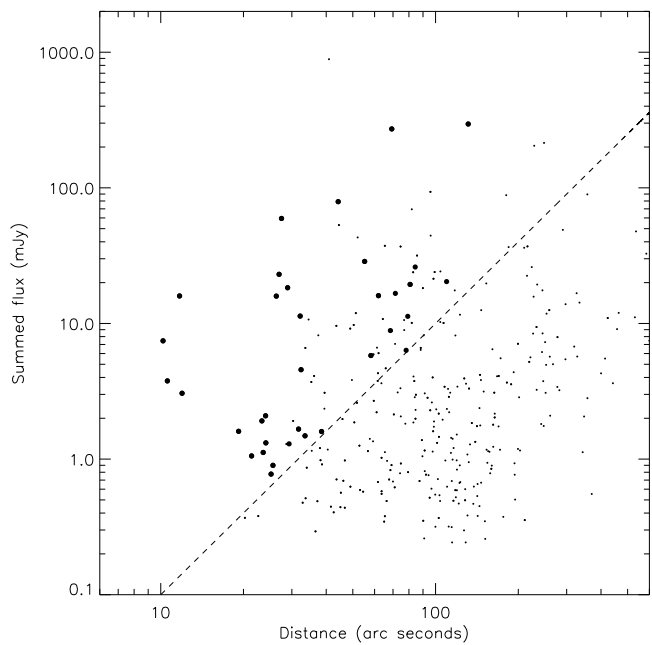

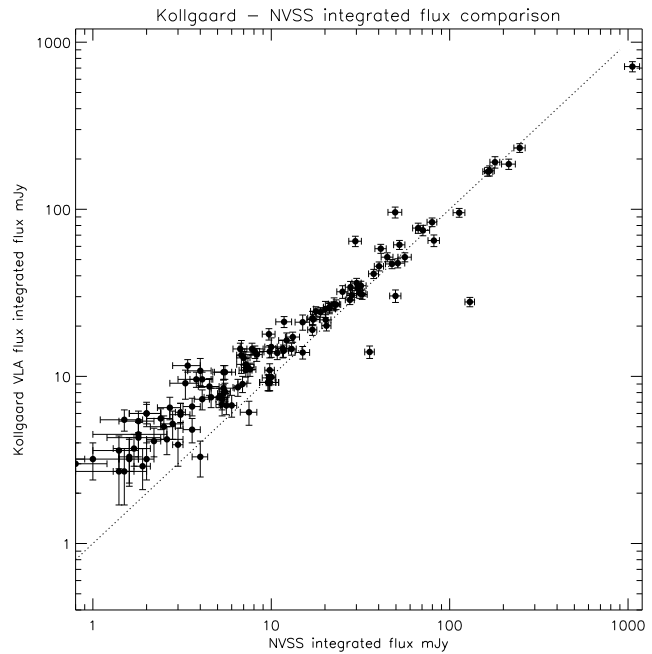

**Fig. 6.** This figure shows the sum of the flux densities of the nearest neighbours between sources from the detection catalogue. Following Magliocchetti et al. (1998) near neighbour pairs to the left of the line are considered as possible double sources. The double sources can be further constrained by adding the second constraint that the fluxes of the two components $f_1$ and $f_2$ should be in the range $0.25 \leq f_1/f_2 \leq 4$, and those sources satisfying this criterion are shown as bold circles in the figure, in other words the multiple component sources, with unhighlighted points representing single component sources.

**Fig. 7.** Cross correlation of the integrated fluxes from the Kollgaard et al. (1994) and the NVSS surveys, with 2 $\sigma$ error bars shown. The fluxes reported by Kollgaard et al. (1994) appear to be systematically higher than the total fluxes reported in the NVSS survey, below $\sim$ 20 mJy.

The positions listed in the Table 1 are those estimated from the external calibration sources and are internally consistent within Table Table 1. Further discussion of the positional alignment to the optical and infrared reference planes will be given in the second paper of this series.

### 5.9. Noise flux accuracy

The accuracy of flux estimates in radio interferometer data has been discussed by a number of authors, for example Rengelink et al. (1997). The accuracy of flux recovery with specific reference to the *SFIND* technique adopted for this paper has been presented in Hopkins et al. (2003), and will not be repeated in detail here. However, for completeness, we will repeat the Hopkins et al. (2003) equations using the terminology in the present paper, which reduce to those presented by Rengelink et al. (1997). For point sources, Hopkins et al. (2003) show that the total relative uncertainty in the integrated flux density is given by:

$$\frac{\sigma_{S_{total}}}{S_{total}} = \sqrt{2.5\frac{\sigma^2}{S_{total}^2} + 0.01^2} \qquad (4)$$

The reader is referred to the papers listed above for more detailed analysis of this, where the treatments of both papers reduce to similar relationships for both extended and for point sources.

### 5.10. Comparison with the VLA flux density scaling

We initially compared the WSRT radio fluxes with those reported in the Kollgaard et al. (1994) paper. This showed that the *WSRT* fluxes below $\sim$ 10 mJy appear to be too bright, initially giving some concern about the calibration efficacy of the *WSRT* data compared to Kollgaard's survey. However, from comparison between the integrated fluxes of the Kollgaard et al. (1994) survey and the NVSS catalogue, it appears that the former study may systematically overestimate the radio fluxes below about 10 mJy, as shown in Fig 7.

It therefore appears that the Kollgaard et al. (1994) fluxes may be unreliable in the flux range appropriate to the *WSRT* survey, and that we should instead rely on NVSS fluxes to assess the efficacy of the present data.

To consider whether the differential source plots are affected by an incorrectly applied completeness correction, or are truly representative of astrophysical the source populations, such as an under-density of sources, or showing evidence for cosmic variance or clustering, we carried out a further series of tests. Firstly, a comparison was made between the fluxes of sources in common between the *VLA* and *WSRT* observations is shown in Fig 8.

To check the efficacy of the *WSRT NEP* data, the flux densities of components measured in the *WSRT* survey were compared with those reported by Kollgaard et al. (1994), as shown in in Fig 8.

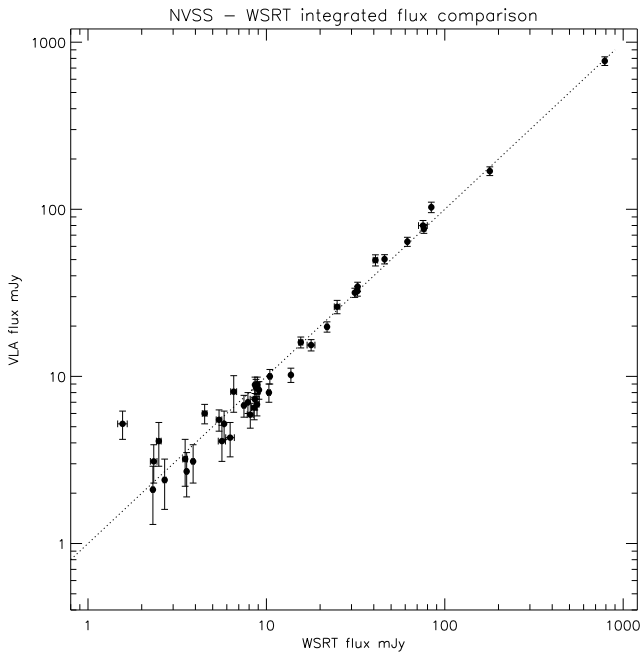

**Fig. 8.** Comparison of the integrated fluxes of isolated radio sources in common between the *NVSS* survey (45″ beam) and the present *WSRT* data lying within one *WSRT* pixel (4″) of each other. The error bars show the ±2 σ uncertainties. Despite the large difference in the beam sizes, the data show good correlation with each other given the errors for fluxes. In the *WSRT* Deep Boötes field, de Vries et al. (2002) have made a detailed comparison of *NVSS* and *WSRT* fluxes, finding their *WSRT* data systematically lower at fainter flux levels, presumably due to a combination of Malmquist and clean biases close to the *NVSS* limit. It is notable that several of the data points in the present survey also show evidence for this trend. A similar effect has also been reported by comparing *ATCA* and *NVSS* observations by Prandoni et al. (2000b).

The two plots show reasonably good agreement between the two independent data sets, with the *VLA* peak fluxes of Kollgaard et al. (1994) systematically lying a little above the *WSRT* peak fluxes above ∼ 50 mJy, but being consistent with the *WSRT* fluxes below that. For the integrated fluxes, there is some evidence that the *WSRT* fluxes are systematically higher than the *VLA* fluxes by ∼ 2–30% at fluxes below 20 mJy, although consistent at higher flux densities. It is difficult to directly compare these, because of the different observational characteristics, and data reduction steps, and there is no *a priori* reason to favour one calibration over another. However, this test does show that at least at the level of a few tens of percent, and down to ∼ 1 mJy, the two data sets are broadly consistent with each other. It is however difficult to know whether or not this holds at lower flux densities, because of the lack of the lower sensitivity of the *VLA* survey, or due to the slightly elliptical beams noted for some parts of the *NEP* field by Kollgaard et al. (1994), which makes direct comparison difficult, differing *uv* coverage, or due to intrinsic radio source variability. It is also notable that Becker et al. (1995) report that one effect of *CLEAN* bias on *VLA* observations is to reduce the flux. On the basis of this, there appears to be no reason to suspect that the measured *WSRT* integrated fluxes lead to the apparent deficit of sub-mJy sources suggested in Figure 11.

A search was then made to count how many sources were detected by the Kollgaard et al. (1994) *VLA* survey to a given flux level in comparison to those detected in the *WSRT* survey. Restricting this analysis to the central 0.5 degree diameter area of the *WSRT* mosaic (where the *rms* peak flux is below ∼ 30 μJy per beam, the *WSRT* observations recover 307 sources with fluxes above 2 mJy, whereas the *VLA* catalogue contains 53 sources, and an unpublished 610 MHz GMRT image (Sirothia et al. *in preparation*) recovers about 312 sources (after making approximate flux scaling corrections assuming that the flux to first order follows a $\nu^{0.75}$ relationship). It therefore appears that there are some unexplained discrepancies between the *VLA* survey of Kollgaard and the *WSRT* results, although in term of raw source numbers, the *WSRT* and GMRT data appear to be more consistent, particularly bearing in mind the approximations assumed about spectral indices. Despite these apparent differences, the check carried out and presented here provide no evidence to support the presence of a systematic bias to the *WSRT* differential number counts presented in Figures 10 and 11, and in the absence of further reasons to be concerned about the *WSRT* counts, it will be assumed that the differences shown in Fig 11 are most likely due to cosmic variance.

Assuming that faint radio sources have the same correlation length as mJy sources from the *FIRST* and *NVSS* surveys ($r \sim 5$ Mpc; Overzier et al. 2003), and that they sample the redshift interval $z = 1 \pm 0.5$, the *rms* uncertainty to the differential source counts in a given flux bin from cosmic variance is estimated to be ∼ 9 per cent (Peacock & Dodds 1994, equation 3 of Somerville et al. 2004, Simpson et al. 2006), which is comparable with the error spread seen at the lower flux levels.

### 5.11. Summary of flux density corrections for systematic effects

As discussed in the previous subsections, various systematic effects have been taken into account to estimate the *WSRT* flux densities, including the clean bias and the bandwidth smearing effects. The corrected flux densities reported in Table 1, ($S_{corr}$) have been corrected for the various effects described as follows (following Prandoni et al. 2000b):

$$S_{corr} = \frac{S_{meas}}{k \left[ a \log \left( \frac{S_{meas}}{\sigma} \right) + b \right]} \quad (5)$$

where $S_{meas}$ is the flux actually measured in the *WSRT* images (reported in the source catalogue in this paper). The parameter $k$ represents the smearing correction. This has a value of 1 (i.e. no correction) when the equation is applied to integrated flux densities and ≤ 1 when dealing with peak flux densities. From Figure 5 we estimate $k = 0.92$ (i.e. an 8% smearing effect which redistributes flux to reduce the *peak* fluxes.).

The clean bias correction is taken into account by the term in the square brackets. As discussed by Prandoni et al. (2000b) the importance of the clean bias effect varies across mosaiced images depending on the average number of clean components. For the present data the average number of clean components was 2121, and on the basis of the simulations reported in Section 5.3 we adopt (a,b) = (0.07,0.82). Applying Equation 5 correctly leads to a good correlation between the *WSRT* and *VLA* fluxes shown in

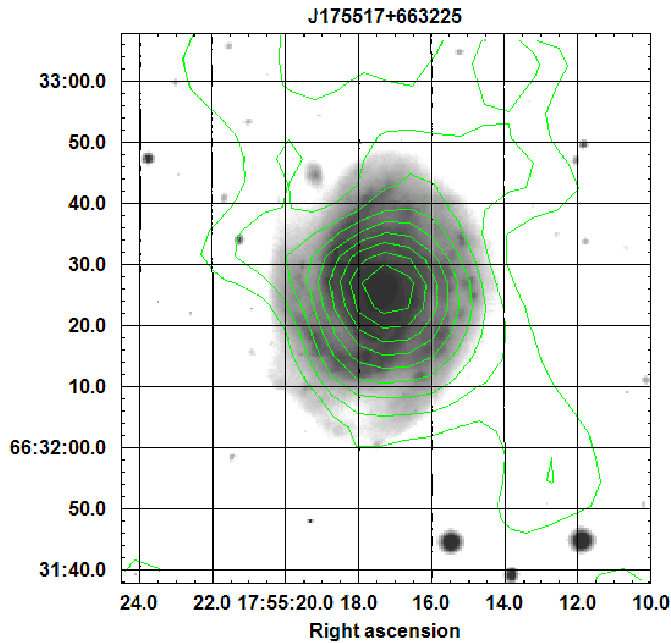
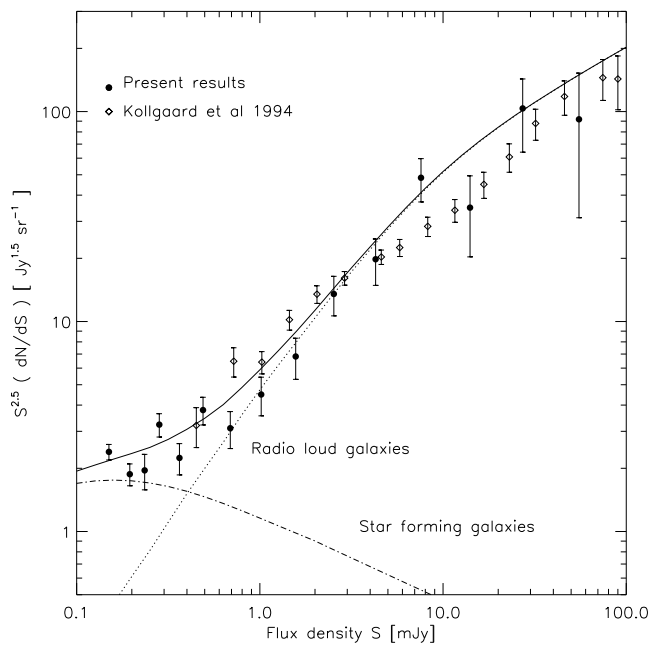

Fig. 9. Overlay between the *WSRT* radio contours and the CFHT R-band image for NEP175517+663225 which is coincident with the 15th magnitude spiral galaxy CGCG 322-021 (from the NED Extragalactic database) at redshift $z = 0.0267$. The *WSRT* contours increase linearly to a peak flux of 0.5 mJy beam$^{-1}$.

Figure 8 (where the *WSRT* fluxes were corrected using these parameters).

## 6. Comparison with other observations

Cross identification with optical data obtained from deep 3.6 metre *CFHT* MEGACAM imaging and IR images from the AKARI survey will be included in the next of this series of papers on the NEP Deep Field. An indicative overlay between the radio and $r'$-band images for the spiral galaxy CGCG 322-021 is shown in Figure 9. Under seeing conditions of 0.87″ the limiting magnitude in this filter was ∼ 23.5. Further details of the data collection and reduction has been given by Hwang et al. (2007).

## 7. Differential Counts

In Figure 10 the differential radio source counts are shown from the *NEP* field, normalised to a static Euclidean Universe (d$N$/d$S$ $S^{2.5}$ (sr$^{-1}$mJy$^{1.5}$)). These source counts are broadly consistent with previous results at 1.4 GHz (e.g. the compilation of Windhorst et al. (1993), the PHOENIX Deep Survey (Hopkins et al. 2003), and the shallow *NEP* survey of Kollgaard et al. 1994).

The data from Fig 10 are given in Table 2, where the flux bins and mean flux for each of the bin centres are listed in columns (1 and 2), the number of sources corrected for clean and resolution bias a discussed in Appendix A are shown in column (3), and the number of sources corrected for the area coverage and double sources in Column 4 [note that because of these correction factors, $N_c$ may be less than $N_0$], and in column (5) we show the differential source counts and the error. The relationship for calculating the numbers in column 5 is the same as that presented by Kollgaard et al. (1994).

Fig. 10. Differential counts determined from the AKARI *NEP* 20 cm deep field. The *WSRT* data points are shown as filled circles, and the diamonds are show the results from the shallow *VLA* survey by Kollgaard et al. (1994). The data from the two surveys are in reasonable accord, although there are some small differences in the flux scales which are believed to be due to two reasons: a) this the same range of flux values where some small differences in the *VLA*–*WSRT* calibration noted in Figure 8, b) this is the region where the correction for double sources most affects the source counts, and it is not clear whether such a correction was made in the *VLA* analysis. The *VLA* data have smaller error bars at the higher flux values because of the lower number of bright sources in the smaller area of the present survey. The model fit to the source counts is shown as the solid line with the composite radio loud and star-forming galaxy populations plotted as dotted and dash-dot lines respectively.

To model the observed source counts a two component model was used that was made of a classical bright radio loud population and a fainter star-forming population. It is well established that classical bright radio galaxies require strong evolution in order to fit the observed source counts at radio wavelengths (Longair 1966, Rowan-Robinson 1970). The source counts above 10 mJy are dominated by giant radio galaxies and QSOs (powered by accretion onto black holes, commonly joined together in the literature under the generic term AGN). Radio loud sources dominate the source counts down to levels of ∼1 mJy, however, at the sub-mJy level the normalised source counts flatten as a new population of faint radio sources emerge (Windhorst et al. 1985). The dominance of starburst galaxies in the sub-mJy population is already well established (Gruppioni et al. 2008), where the number of blue galaxies with star-forming spectral signatures is seen to increase strongly. Rowan-Robinson et al. (1993), Hopkins et al. (1998), and others have concluded that the source counts at these faintest levels require two populations, AGNs and starburst galaxies. This latter population can best be modelled as a dusty star-forming population, under the assumption that they are the higher redshift analogues of the IRAS star-forming population (Rowan-Robinson et al. 1993, Pearson & Rowan-Robinson 1996). In this scenario, the radio emission originates from the non-thermal synchrotron emission from relativistic electrons accelerated by supernovae remnants in the host galaxies.

Table 2. 20 cm differential source counts for the $WSRT-AKARI-NEP$ survey

| Flux bin mJy (1) | Mean Flux mJy (2) | $N_0$ (3) | $N_c$ (4) | $dN/dS$ $sr^{-1}$ $Jy^{1.5}$ (5) |
|---|---|---|---|---|
| 0.130–0.170 | 0.150 | 31 | 138.3 | 2.39±0.20 |
| 0.170–0.220 | 0.195 | 31 | 70.1 | 1.87±0.22 |
| 0.220–0.250 | 0.235 | 14 | 27.02 | 1.95±0.38 |
| 0.250–0.315 | 0.283 | 44 | 61.87 | 3.23±0.41 |
| 0.315–0.413 | 0.364 | 33 | 34.60 | 2.24±0.38 |
| 0.413–0.566 | 0.489 | 48 | 43.95 | 3.78±0.57 |
| 0.566–0.813 | 0.689 | 41 | 24.98 | 3.10±0.62 |
| 0.813–1.221 | 1.017 | 40 | 22.89 | 4.49±0.94 |
| 1.221–1.917 | 1.569 | 37 | 20.31 | 6.82±1.51 |
| 1.917–3.151 | 2.534 | 38 | 21.90 | 13.53±2.89 |
| 3.151–5.416 | 4.283 | 29 | 16.10 | 19.80±4.93 |
| 5.416–9.742 | 7.597 | 34 | 18.38 | 48.39±11.29 |
| 9.742–18.33 | 14.04 | 14 | 5.74 | 34.87±14.55 |
| 18.33–36.08 | 27.21 | 10 | 6.88 | 103.53±39.47 |
| 36.08–74.32 | 55.20 | 5 | 2.29 | 91.86±60.66 |

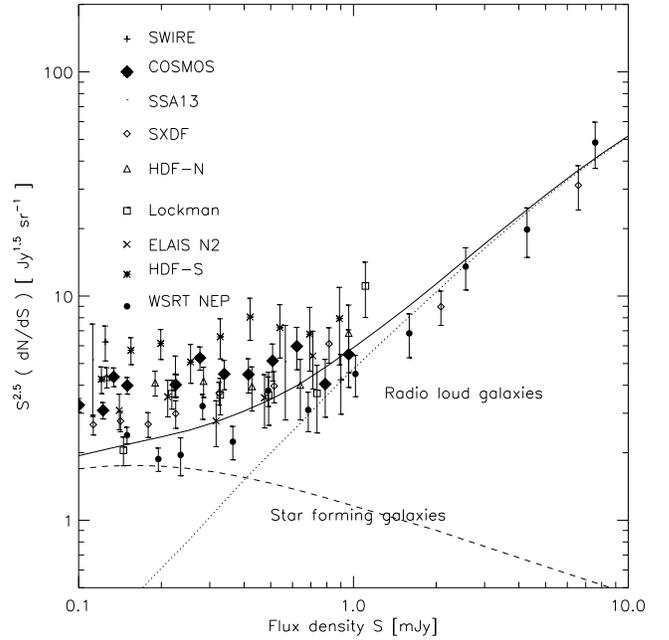

**Fig. 11.** A compilation of the differential source counts of a number of deep 20 cm radio surveys taken from: SWIRE Owen & Morrison 2008; COSMOS Bondi et al. 2008; SSA13 Fomalont et al. 2006; SXDF Simpson et al. 2006; HDF-N, LOCKMAN and ELAIS N2 Biggs & Ivison 2006, HDF-S Huynh et al. 2005. The solid curve is the best fit to the present data taken as described in Fig 10. There are however differences in the instrumental and systematic corrections that have been made for the different survey results shown here (see detailed discussion by Prandoni et al. 2000b), that make quantitative comparison at the faintest flux levels somewhat uncertain.

To represent the radio loud population the luminosity function of Dunlop & Peacock (1990) was used to model the local space density with an assumption that the population evolves in luminosity with increasing redshift. The assumed luminosity evolution then follows a power law with redshift of $(1+z)^{3.1}$, broadly consistent with both optically and X-ray selected quasars (Boyle et al. 1987). The spectrum of the radio loud quasar population was modelled from Elvis, Lockman & Fassnacht (1994), assuming a steep radio spectrum source of ($S_\nu \propto \nu^{-\alpha}$, $\alpha$=1).

To model the faint sub-mJy population the $IRAS$ 60 $\mu$m luminosity function of Saunders et al. (2000) was used as a starting point, with the parameters for the star-forming population, segregated by warmer 100 $\mu$m / 60 $\mu$m $IRAS$ colour, given in Pearson (2005, 2010). To convert the infrared luminosity function to radio wavelengths, the well known correlation between the 60 $\mu$m far-IR and radio flux emission of $S_{60\,\mu m} = 90S(1.4\text{ GHz})$ (Helou, Soifer & Rowan-Robinson 1985, Yun, Reddy & Condon (2001), Appleton et al. (2004), White et al. (2009) was assumed. To model the star-forming population the spectral template of the archetypical starburst galaxy of M82 from the models of Efstathiou, Rowan-Robinson & Siebenmorgen (2000) was adopted. The radio and far-infrared fluxes are correlated due to the presence of hot OB stars in giant molecular clouds that heat the surrounding dust producing the infrared emission. These stars subsequently end their lives as supernovae with the radio emission powered by the synchrotron emission from their remnants. The radio spectrum is characterised by a power law of ($S_\nu \propto \nu^{-\alpha}$, $\alpha$=0.8).

It was assumed that the star-forming population evolves in luminosity as a power law $\propto (1+z)^{3.0}$. This infrared representation of the star-forming population was preferred over using the radio luminosity function directly, since it creates a phenomenological link between the radio emission and the infrared which is responsible for the bulk of the emission in the star-forming population. Note that Huynh et al. (2005) used the radio luminosity function of Condon et al. (2002) and derived a best fitting evolution parameterisation $\propto (1+z)^{2.7}$, slightly lower than the work presented here although the values are broadly consistent and differences can be due to the assumed SED and luminosity function. Hopkins (2004) and Hopkins et al. (1998) used radio and infrared luminosity functions respectively obtaining evolution in the sub-mJy population $\propto (1+z)^{2.7}$ and $\propto (1+z)^{3.3}$. It does however appear that the counts measured in this study lie at the lower end of the emerging picture on excess sub-mJy radio counts, as shown in Fig 11.

## 8. Conclusions

A deep radio survey has been made of an $\sim 1.7$ square degree area around the North Ecliptic Pole field using the $WSRT$ at 20 cm wavelength. The maximum sensitivity of the survey was 21 $\mu$Jy beam$^{-1}$, with a synthesised beam of $17.0 \times 15.5''$. The analysis methodology was carefully chosen to mitigate the various effects that can affect the efficacy of radio synthesis array observations, resulting in a final catalogue of 462 radio emitting sources, with the faintest integrated fluxes at about the 100 $\mu$Jy level. The differential source counts calculated from the $WSRT$ data show a pronounced excess for sources fainter than $\sim 1$ mJy, consistent with a population of faint star forming galaxies. Comparison between the Kollgaard et al. (1994) survey and the NVSS catalogue shows a systematic difference in this flux range, suggesting that one or the other may suffer from a slight mis-calibration. The present *WSRT* catalogue of radio sources will form the basis for two further papers reporting cross correlation against extant AKARI and deep optical imaging. A further paper reporting the radio spectral indices of the sources utilising *GMRT* data will be reported elsewhere.

## 9. Acknowledgements

This work is based on observations with AKARI, a *JAXA* project with the participation of ESA. We thank Andrew Hopkins for helping us to understand the operation of *SFIND*, Niruj Mohan for discussions on source detection algorithms, and Sandeep Sirothia for discussions about the GMRT 610 MHz NEP survey. We also express our thanks to The Netherlands Institute for Radio Astronomy, *ASTRON*, for the substantial allocation of observing time; the staff of the Westerbork Observatory for technical support; and the UK Science and Technology Facilities Council, *STFC* and its forerunner, *PPARC*, for manpower and travel support. The UK-Japan AKARI Consortium has also received funding awards from the Sasakawa Foundation, The British Council, and the DAIWA Foundation, which facilitated travel and exchange activities, and for which we are very grateful. T.G. acknowledges financial support from the Japan Society for the Promotion of Science (JSPS) through JSPS Research Fellowships for Young Scientists. MI was supported by the Korea Science and Engineering Foundation(KOSEF) grant No. 2009-0063616, funded by the Korea government(MEST).

## References

Affer, L., Micela, G., Morel, T. 2008, A&A, 483, 801
Appleton, P. N., Fadda, D. T., Marleau, F. R. et al. 2004, ApJS, 154, 147
Aussel, H., Coia, D., Mazzei, P. et al. 2000, A&AS, 141, 257
Becker, R.H., White, R.L. & Helfand, D.J 1995, ApJ, 450, 559
Bennett, C. L., Banday, A. J., Gorski, K. M. et al. 1996, ApJ, 464, 1
Biggs, A.D., Ivison, R.J. 2006, MNRAS, 371, 963
Bondi, M., Ciliegi, P., Zamorani, G. et al. 2003, A&AS, 403, 857
Bondi, M., Ciliegi, P., Schinnerer, E. et al. 2008, ApJ, 681, 1129
Boyle B.J., Fong R., Shanks T. et al. 1987, MNRAS, 227, 717
Branchesi, M., Gioia, I. M., Fanti, C. et al. 2006, A&A, 446, 97
Bridle, A.H., & Schwab, F.R. 1989, in Synthesis Imaging in Radio Astronomy, ed R.A. Perley, F.R Schwab & A.H.Bridle (ASP Conf. Ser. 6),247
Brinkmann, W., Chester, M., Kollgaard, R., et al. 1999, A&AS, 134, 221; and VizieR On-line Data Catalog: J/A+AS/134/221
Condon, J. J., Broderick, J.J. 1985, AJ, 90, 2540
Condon, J. J., Broderick, J.J. 1986, AJ, 91, 1051
Condon, J. J., Cotton, W. D., Greisen, E. W. et al. 1998, AJ, 115, 1693
Condon, J.J., Cotton, W.D., Broderick, J.J. 2002, AJ, 124, 675
Coppin, K., Chapin, E. L., Mortier, A. M. J. et al. 2006, MNRAS, 372, 1621
Cotton, W.H. 1989, in Synthesis Imaging in Radio Astronomy, ed R.A. Perley, F.R Schwab & A.H.Bridle (ASP Conf. Ser. 6),233
de Vries, W.H., Morganti, R., Röttgering, H.J.A. et al. 2002, AJ, 1231, 1784
Dunlop J.S., Peacock J.A., 1990, MNRAS, 247, 19
Efstathiou A., Rowan-Robinson M., Siebenmorgen R., 2000, MNRAS, 313, 734
Elvis M., Lockman, F.J., Fassnacht, C. 1994, ApJS, 95, 413
Fomalont, E. B., Kellermann, K. I., Cowie, L. L. et al. 2006, ApJS, 167, 103.
Gaidos, E.J., Magnier, E.A., Schechter, P.L. et al. 1993, PASP, 105, 1294
Garn, T., Green, D.A., Riley, J.M., et al. 2000, MNRAS, 383, 75.
Garrett, M.A., de Bruyn, A.G., M. Giroletti, M. et al. 2000, A&A, 361, 41.
Gioia, I.M., Henry, J.P., Mullis, C.R. et al. 2003, ApJS, 149, 29
Gioia, I.M., Wolter, A., Mullis, C.R. et al. 2004, A&A, 428, 867
Goto, T., Hanami, H., Im, M. et al. 2008, PASJ, 60, 531
Gruppioni, C., Pozzi, F., Polletta, M. et al. 2008, MNRAS, 684, 136
Hacking, P., Condon, J.J. & Houck, J.R. 1987, ApJ, 316, L15
Helou G., Soifer T.T., Rowan-Robinson M. 1985, ApJ, 298, 7
Henry, J. P., Gioia, I. M., Mullis, C. R. et al. 2001, ApJ, 553, 109
Henry, J.P., Mullis, C.R., Voges, W. et al. 2006, ApJS, 162, 304
Hopkins A.M., Mobasher B., Cram L. et al. 1998, MNRAS, 296, 839
Hopkins, A., Afonso, J., Cram, L. et al. 1999, ApJ., 519, L59
Hopkins, A.M., Miller, C.J., Connolly, A.J. et al. 2002, AJ, 123,1086
Hopkins A.M., Afonso J., Chan B. et al. 2003, AJ, 125, 465
Hopkins, A. M. 2004, ApJ, 615, 209
Huynh M.T., Jackson C.A., Norris R.P. et al. , 2005, AJ, 130, 1388
Hwang, N., Lee, H-M., Im, M. et al. , 2007, ApJS, 172, 583
Kaper H.G., Smiths D.W., Schwartz U. et al. ., 1966, BAN 18, 465
Kollgaard, R.I., Brinkmann, W., Chester, M.M. et al. 1994, ApJS, 93, 145
Kümmel, M.W. & Wagner, S.J. 2000, A&A, 353, 867
Kümmel, M.W. & Wagner, S.J. 2001, A&A, 370, 384
Lee, H.M., Kim, S.J., Im, M. et al. 2009, astro-ph/0901.3256
Loiseau, N., Reich, W., Wielebinski, R., et al. 1988, A&AS, 75,67
Longair M.S., 1966, MNRAS, 133, 421
Magliocchetti, M., Maddox, S.J., Lahav, O. et al. 1998. MNRAS, 300, 257
Matsuhara, H., Wada, T., Matsuura, S. et al. 2006, PASJ, 58, 673
Micela, G., Affer, L., Favata, F. et al. 2007, A&A, 461, 977
Morganti, R., Garrett, M. A., Chapman, S. et al. 2004. A&A, 424, 371.
Moss, D., Seymour, N., McHardy, I. et al. 2007, MNRAS, 378, 995
Mullis, C. R., Henry, J. P., Gioia, I. M. et al. 2001, ApJ, 553, 115
Mullis, C.R., McNamara, B.R., Quintana, H. et al. 2003, ApJ, 594, 154
Owen, F.R. & Morrison, G.E.. 2008. AJ, 136, 1889.
Oosterbaan, C.E. 1978, A&A, 69, 235
Overzier R. A., Rottgering H. J. A., Rengelink R. B. et al. 2003, A&A, 405, 53
Pearson C.P. & Rowan-Robinson M., 1996, MNRAS, 283, 174
Pearson C.P. 2005, MNRAS 358, 1417
Pearson C.P. 2010, in preparation
Peacock J. A., Dodds S. J., 1994, MNRAS, 267, 1020
Prandoni, I., Gregorini, L., Parma, P. et al. 2000a, A&A Suppl, 146, 31
Prandoni, I., Gregorini, L., Parma, P. et al. 2000b, A&A Suppl, 146, 41
Prandoni, I., Gregorini, L., Parma, P. et al. 2001, A&A, 365, 392
Prandoni I., Parma P., Wieringa M. H., et al. 2006, A&A, 457, 517
Pretorius, M. L., Knigge, C. & O'Donoghue, D. 2007. MNRAS, 382, 1279
Rengelink, R.B., Tang, Y., de Bruyn, A.G. et al. 1997, A&A Suppl., 124, 259
Rowan-Robinson M., 1970, MNRAS, 149, 365
Rowan-Robinson M., Benn C.R., Lawrence A. et al. 1993, MNRAS, 263, 192
Sault R.J., Teuben P.J., Wright M.C.H., 1995, *Astronomical Data Analysis Software and Systems IV*, ed. R. Shaw, H.E. Payne, J.J.E. Hayes, ASP Conf. Ser., 77, 433-436
Saunders W., Sutherland, W.J., Maddox, S.J. et al. 2000, MNRAS, 317, 55
Sedgwick,C., Serjeant,S., Sirothia,S., Pal, S. et al. . 2010. To appear in Astronomical Society of the Pacific Conference Series, vole 418, Editors Onaka, T., White, G.J., Nakagawa, T. and Yamamura, I.
Simpson, C., Martinez-Sansigre, A., Rawlings, S. et al. 2006. MNRAS, 372, 741
Somerville R. S., Lee K., Ferguson H. C.et al. 2004, ApJ, 600, L171
Stickel, M., Bogun, S., Lemke, D. et al. 1998, A&A, 336, 116
Voges, W., Henry, J.P., Briel, U.G. et al. 2001, ApJ, 553, 119
Wada, T., Matsuhara, H., Oyabu, S. et al. 2008, PASJ, 60,517
White, R. L., Becker, R. H., Helfand, D. J. et al. 1997, ApJ, 475, 479
Windhorst, R.A., Miley G.K., Oweb, F.N. et al. 1985, ApJ, 289, 494
Windhorst R.A., Fomalont E.B., Partridge R.B. et al. 1993, ApJ, 405, 498
Yun, M.S., Reddy, N.A., Condon, J.J. 2001, ApJ, 554, 803

Zickgraf, F., Thiering, I., Krautter, J. et al. 1997, A&AS, 123, 103

Table 1.

| Running number | Source name | RA hh:mm:ss.s | DEC dd:mm:ss.s | δRA " | δDEC " | $S_{peak}$ mJy beam$^{-1}$ | $S_{peak}$ error mJy beam$^{-1}$ | $S_{total}$ mJy | $S_{total}$ error mJy | $\theta_{maj}$ " | $\theta_{min}$ " | PA ° |
|---|---|---|---|---|---|---|---|---|---|---|---|---|
| (1) | (2) | (3) | (4) | (5) | (6) | (7) | (8) | (9) | (10) | (11) | (12) | (13) |
| 1 | NEP175121+663645 | 17:51:21.7 | +66:36:45.1 | 1.39 | 0.22 | 1.219 | 0.270 | 1.932 | 0.295 | 28.0 | | -81.6 |
| 2 | NEP175140+665038 | 17:51:41.0 | +66:50:38.1 | 0.01 | 0.01 | 2.464 | 0.364 | 1.461 | 0.367 | | | |
| 3 | NEP175147+663124 | 17:51:47.8 | +66:31:24.1 | 0.05 | 0.04 | 1.138 | 0.177 | 0.829 | 0.182 | | | |
| 4 | NEP175214+665054 | 17:52:14.7 | +66:50:54.2 | 0.05 | 0.11 | 2.615 | 0.257 | 5.297 | 0.288 | 24.2 | 13.0 | -0.7 |
| 5 | NEP175231+662738 | 17:52:31.9 | +66:27:38.3 | 0.38 | 0.28 | 0.633 | 0.148 | 1.181 | 0.158 | 20.6 | 14.8 | -54.6 |
| 6 | NEP175248+662713 | 17:52:48.6 | +66:27:13.9 | 0.01 | 0.01 | 0.886 | 0.102 | 0.885 | 0.103 | | | |
| 7 | NEP175254+663144 | 17:52:54.5 | +66:31:44.1 | 0.14 | 0.08 | 2.774 | 0.131 | 7.340 | 0.184 | 29.7 | 17.4 | -86.5 |
| 8 | NEP175256+663148 | 17:52:56.4 | +66:31:48.4 | 0.74 | 0.10 | 2.078 | 0.131 | 7.709 | 0.204 | 59.6 | 9.7 | 85.5 |
| 9 | NEP175305+663929 | 17:53:06.0 | +66:39:29.8 | 0.02 | 0.01 | 1.708 | 0.121 | 2.216 | 0.131 | | | |
| 10 | NEP175307+663213 | 17:53:07.6 | +66:32:13.2 | 0.02 | 0.02 | 3.735 | 0.131 | 8.035 | 0.202 | 21.1 | 17.4 | 8.4 |
| 11 | NEP175313+661949 | 17:53:13.9 | +66:19:49.7 | 0.00 | 0.00 | 6.197 | 0.148 | 8.574 | 0.186 | | | |
| 12 | NEP175321+661249 | 17:53:21.5 | +66:12:49.1 | 0.04 | 0.03 | 1.403 | 0.149 | 1.673 | 0.153 | | | |
| 13 | NEP175330+662831 | 17:53:30.0 | +66:28:31.5 | 0.00 | 0.00 | 1.056 | 0.086 | 1.248 | 0.090 | | | |
| 14 | NEP175331+662726 | 17:53:31.3 | +66:27:26.6 | 0.00 | 0.00 | 26.929 | 0.086 | 32.232 | 0.147 | | | |
| 15 | NEP175334+665105 | 17:53:34.8 | +66:51:05.2 | 0.01 | 0.00 | 1.503 | 0.148 | 2.112 | 0.149 | | | |
| 16 | NEP175335+663547 | 17:53:35.1 | +66:35:47.9 | 0.00 | 0.00 | 152.400 | 0.224 | 180.740 | 0.585 | | | |
| 17 | NEP175336+663438 | 17:53:36.0 | +66:34:38.8 | 0.00 | 0.00 | 41.663 | 0.147 | 61.508 | 0.322 | | | |
| 18 | NEP175338+663135 | 17:53:38.4 | +66:31:35.4 | 0.02 | 0.03 | 5.419 | 0.090 | 8.668 | 0.267 | | | |
| 19 | NEP175338+665105 | 17:53:38.8 | +66:51:05.2 | 0.00 | 0.00 | 1.503 | 0.148 | 2.321 | 0.380 | | | |
| 20 | NEP175339+663319 | 17:53:39.3 | +66:33:19.3 | 0.01 | 0.01 | 1.197 | 0.147 | 1.120 | 0.148 | | | |

Table 1. −continued

| Running number | Source name | RA hh:mm:ss.s | DEC dd:mm:ss.s | δRA " | δDEC " | $S_{peak}$ mJy beam$^{-1}$ | $S_{peak}$ error mJy beam$^{-1}$ | $S_{total}$ mJy | $S_{total}$ error mJy | $\theta_{maj}$ " | $\theta_{min}$ " | PA ° |
|---|---|---|---|---|---|---|---|---|---|---|---|---|
| (1) | (2) | (3) | (4) | (5) | (6) | (7) | (8) | (9) | (10) | (11) | (12) | (13) |
| 21 | NEP175348+663922 | 17:53:48.3 | +66:39:22.7 | 0.00 | 0.00 | 0.736 | 0.081 | 0.715 | 0.084 | | | |
| 22 | NEP175357+670348 | 17:53:57.8 | +67:03:48.4 | 0.05 | 0.03 | 3.387 | 0.327 | 6.131 | 0.350 | 25.2 | 9.5 | -58.6 |
| 23 | NEP175358+664929 | 17:53:58.6 | +66:49:29.6 | 0.07 | 0.05 | 1.111 | 0.101 | 1.738 | 0.114 | 18.6 | 10.7 | -64.5 |
| 24 | NEP175410+664620 | 17:54:10.2 | +66:46:20.3 | 0.03 | 0.06 | 0.652 | 0.084 | 0.987 | 0.087 | 24.2 | | -5.9 |
| 25 | NEP175412+663401 | 17:54:12.8 | +66:34:02.0 | 0.03 | 0.02 | 0.410 | 0.063 | 0.476 | 0.065 | | | |
| 26 | NEP175419+660245 | 17:54:19.9 | +66:02:45.7 | 0.11 | 0.14 | 1.839 | 0.360 | 5.392 | 0.386 | 28.6 | 22.3 | -13.0 |
| 27 | NEP175420+663059 | 17:54:20.4 | +66:30:59.1 | 0.00 | 0.00 | 2.681 | 0.048 | 2.822 | 0.055 | | | |
| 28 | NEP175437+662248 | 17:54:37.3 | +66:22:48.6 | 0.01 | 0.01 | 0.722 | 0.072 | 0.826 | 0.073 | | | |
| 29 | NEP175438+662319 | 17:54:38.8 | +66:23:19.0 | 0.01 | 0.01 | 0.708 | 0.072 | 0.740 | 0.074 | | | |
| 30 | NEP175445+664823 | 17:54:45.5 | +66:48:24.0 | 0.03 | 0.03 | 0.626 | 0.064 | 0.737 | 0.067 | | | |
| 31 | NEP175445+661208 | 17:54:46.0 | +66:12:08.7 | 0.01 | 0.01 | 14.286 | 0.091 | 25.459 | 0.455 | 17.6 | 13.6 | 81.2 |
| 32 | NEP175454+663417 | 17:54:54.0 | +66:34:17.6 | 0.06 | 0.04 | 0.227 | 0.038 | 0.171 | 0.039 | | | |
| 33 | NEP175455+665658 | 17:54:55.4 | +66:56:58.3 | 0.00 | 0.00 | 3.637 | 0.124 | 4.334 | 0.141 | | | |
| 34 | NEP175458+661723 | 17:54:58.9 | +66:17:23.7 | 0.01 | 0.01 | 1.453 | 0.070 | 1.546 | 0.079 | | | |
| 35 | NEP175459+661139 | 17:54:59.3 | +66:11:39.9 | 0.02 | 0.03 | 3.273 | 0.085 | 4.427 | 0.155 | | | |
| 36 | NEP170550+663153 | 17:05:50.0 | +66:31:53.2 | 0.01 | 0.01 | 0.390 | 0.051 | 0.421 | 0.051 | | | |
| 37 | NEP175504+663952 | 17:55:04.8 | +66:39:52.7 | 0.00 | 0.00 | 8.001 | 0.055 | 8.472 | 0.064 | | | |
| 38 | NEP175506+661737 | 17:55:06.6 | +66:17:37.7 | 0.00 | 0.00 | 6.142 | 0.070 | 6.925 | 0.092 | | | |
| 39 | NEP175509+663400 | 17:55:09.4 | +66:34:00.0 | 0.00 | 0.00 | 2.115 | 0.038 | 2.455 | 0.045 | | | |
| 40 | NEP175512+663040 | 17:55:12.6 | +66:30:40.1 | 0.04 | 0.06 | 1.657 | 0.051 | 3.415 | 0.083 | 24.3 | 12.7 | -28.5 |

Table 1. – *continued*

| Running number | Source name | RA hh:mm:ss.s | DEC dd:mm:ss.s | δRA " | δDEC " | S$_{peak}$ mJy beam$^{-1}$ | S$_{peak}$ error mJy beam$^{-1}$ | S$_{total}$ mJy | S$_{total}$ error mJy | θ$_{maj}$ " | θ$_{min}$ " | PA ° |
|---|---|---|---|---|---|---|---|---|---|---|---|---|
| (1) | (2) | (3) | (4) | (5) | (6) | (7) | (8) | (9) | (10) | (11) | (12) | (13) |
| 41 | NEP175513+663111 | 17:55:13.5 | +66:31:11.8 | 0.04 | 0.03 | 3.626 | 0.045 | 7.360 | 0.195 | 21.3 | 14.2 | 56.3 |
| 42 | NEP175515+655429 | 17:55:15.2 | +65:54:29.7 | 0.04 | 0.05 | 2.220 | 0.348 | 1.766 | 0.359 | | | |
| 43 | NEP175515+661703 | 17:55:15.8 | +66:17:03.7 | 0.01 | 0.01 | 1.361 | 0.070 | 1.606 | 0.077 | | | |
| 44 | NEP175516+662425 | 17:55:16.9 | +66:24:25.6 | 0.00 | 0.00 | 1.263 | 0.037 | 1.341 | 0.042 | | | |
| 45 | NEP175517+662812 | 17:55:17.0 | +66:28:13.0 | 0.01 | 0.01 | 0.258 | 0.041 | 0.179 | 0.041 | | | |
| 46 | NEP175517+663225 | 17:55:17.1 | +66:32:25.6 | 0.04 | 0.05 | 0.517 | 0.045 | 1.158 | 0.051 | 23.3 | 17.5 | 3.1 |
| 47 | NEP175517+663919 | 17:55:17.4 | +66:39:19.2 | 0.02 | 0.03 | 0.527 | 0.036 | 0.604 | 0.040 | | | |
| 48 | NEP175517+661658 | 17:55:17.4 | +66:16:58.8 | 0.95 | 0.53 | 0.819 | 0.055 | 1.991 | 0.085 | 27.7 | 15.9 | -63.5 |
| 49 | NEP175520+662428 | 17:55:20.9 | +66:24:28.2 | 0.00 | 0.00 | 0.410 | 0.037 | 0.631 | 0.046 | 19.4 | 8.8 | -64.2 |
| 50 | NEP175521+663526 | 17:55:21.7 | +66:35:26.2 | 0.02 | 0.01 | 0.397 | 0.043 | 0.408 | 0.044 | | | |
| 51 | NEP175521+661338 | 17:55:21.9 | +66:13:38.4 | 0.05 | 0.02 | 38.490 | 0.135 | 79.820 | 2.134 | 22.6 | 12.8 | -89.9 |
| 52 | NEP175523+662404 | 17:55:23.8 | +66:24:04.7 | 0.01 | 0.02 | 0.897 | 0.037 | 0.879 | 0.045 | | | |
| 53 | NEP175526+665356 | 17:55:26.8 | +66:53:56.0 | 0.00 | 0.00 | 6.096 | 0.084 | 7.244 | 0.100 | | | |
| 54 | NEP175527+662351 | 17:55:27.0 | +66:23:51.2 | 0.10 | 0.09 | 0.678 | 0.037 | 0.968 | 0.057 | | | |
| 55 | NEP175529+662416 | 17:55:29.6 | +66:24:16.0 | 0.11 | 0.11 | 0.199 | 0.037 | 0.290 | 0.039 | 17.6 | 9.7 | 60.0 |
| 56 | NEP175533+663116 | 17:55:33.2 | +66:31:16.8 | 0.00 | 0.00 | 0.470 | 0.045 | 0.409 | 0.045 | | | |
| 57 | NEP175535+662744 | 17:55:35.8 | +66:27:44.4 | 0.03 | 0.05 | 0.237 | 0.041 | 0.295 | 0.042 | | | |
| 58 | NEP175537+664550 | 17:55:37.7 | +66:45:50.1 | 0.24 | 0.06 | 0.239 | 0.046 | 0.378 | 0.049 | 25.1 | 5.0 | 88.5 |
| 59 | NEP175545+664037 | 17:55:45.3 | +66:40:37.4 | 0.33 | 0.24 | 0.177 | 0.039 | 0.255 | 0.041 | 20.9 | 4.8 | 60.5 |
| 60 | NEP175546+663839 | 17:55:46.2 | +66:38:39.5 | 0.07 | 0.11 | 0.247 | 0.039 | 0.320 | 0.041 | 18.3 | | -30.7 |

Table 1. −continued

| Running number | Source name | RA hh:mm:ss.s | DEC dd:mm:ss.s | δRA " | δDEC " | $S_{peak}$ mJy beam$^{-1}$ | $S_{peak}$ error mJy beam$^{-1}$ | $S_{total}$ mJy | $S_{total}$ error mJy | $θ_{maj}$ " | $θ_{min}$ " | PA ° |
|---|---|---|---|---|---|---|---|---|---|---|---|---|
| (1) | (2) | (3) | (4) | (5) | (6) | (7) | (8) | (9) | (10) | (11) | (12) | (13) |
| 61 | NEP175548+670656 | 17:55:48.1 | +67:06:56.8 | 0.63 | 0.51 | 0.948 | 0.198 | 3.173 | 0.235 | 29.7 | 25.8 | -62.1 |
| 62 | NEP175550+671240 | 17:55:50.4 | +67:12:40.9 | 1.61 | 0.45 | 3.369 | 0.383 | 1.212 | 0.383 | | | |
| 63 | NEP175552+662635 | 17:55:52.2 | +66:26:35.8 | 0.11 | 0.18 | 0.255 | 0.041 | 0.513 | 0.046 | 26.2 | 11.1 | 23.0 |
| 64 | NEP175555+663629 | 17:55:55.6 | +66:36:29.2 | 0.15 | 0.10 | 0.223 | 0.035 | 0.167 | 0.038 | | | |
| 65 | NEP175601+660829 | 17:56:01.1 | +66:08:29.8 | 0.02 | 0.02 | 1.803 | 0.093 | 3.058 | 0.112 | 19.7 | 11.3 | 23.2 |
| 66 | NEP175601+663500 | 17:56:01.9 | +66:35:00.0 | 0.14 | 0.14 | 0.330 | 0.035 | 0.427 | 0.041 | | | |
| 67 | NEP175602+661824 | 17:56:02.1 | +66:18:24.7 | 0.00 | 0.00 | 1.421 | 0.044 | 1.467 | 0.048 | | | |
| 68 | NEP175604+660805 | 17:56:04.7 | +66:08:05.9 | 0.03 | 0.04 | 0.925 | 0.093 | 1.266 | 0.099 | | | |
| 69 | NEP175608+664119 | 17:56:08.2 | +66:41:19.6 | 0.00 | 0.01 | 0.361 | 0.039 | 0.369 | 0.039 | | | |
| 70 | NEP175608+664424 | 17:56:08.7 | +66:44:24.2 | 0.01 | 0.01 | 0.712 | 0.045 | 0.725 | 0.047 | | | |
| 71 | NEP175609+663304 | 17:56:09.6 | +66:33:04.5 | 0.00 | 0.00 | 0.555 | 0.035 | 0.542 | 0.036 | | | |
| 72 | NEP175611+663542 | 17:56:11.5 | +66:35:42.0 | 0.04 | 0.08 | 0.202 | 0.031 | 0.177 | 0.033 | | | |
| 73 | NEP175612+661731 | 17:56:12.7 | +66:17:31.9 | 0.05 | 0.04 | 0.234 | 0.043 | 0.276 | 0.044 | | | |
| 74 | NEP175615+664653 | 17:56:15.8 | +66:46:53.8 | 0.03 | 0.04 | 0.367 | 0.036 | 0.363 | 0.038 | | | |
| 75 | NEP175620+664340 | 17:56:20.8 | +66:43:40.8 | 0.14 | 0.06 | 1.116 | 0.039 | 2.864 | 0.091 | 32.6 | 13.3 | 67.1 |
| 76 | NEP175621+661649 | 17:56:21.7 | +66:16:50.0 | 0.01 | 0.01 | 0.548 | 0.040 | 0.512 | 0.041 | | | |
| 77 | NEP175625+663243 | 17:56:25.2 | +66:32:43.4 | 0.01 | 0.01 | 1.323 | 0.034 | 1.620 | 0.047 | | | |
| 78 | NEP175628+663820 | 17:56:29.0 | +66:38:20.6 | 0.04 | 0.03 | 0.338 | 0.031 | 0.364 | 0.033 | | | |
| 79 | NEP175631+662509 | 17:56:31.8 | +66:25:09.6 | 0.06 | 0.11 | 0.163 | 0.029 | 0.175 | 0.030 | | | |
| 80 | NEP175634+665317 | 17:56:34.3 | +66:53:17.5 | 0.02 | 0.03 | 0.318 | 0.060 | 0.305 | 0.061 | | | |

Table 1. – *continued*

| Running number | Source name | RA hh:mm:ss.s | DEC dd:mm:ss.s | δRA " | δDEC " | S$_{peak}$ mJy beam$^{-1}$ | S$_{peak}$ error mJy beam$^{-1}$ | S$_{total}$ mJy | S$_{total}$ error mJy | θ$_{maj}$ " | θ$_{min}$ " | PA ° |
|---|---|---|---|---|---|---|---|---|---|---|---|---|
| (1) | (2) | (3) | (4) | (5) | (6) | (7) | (8) | (9) | (10) | (11) | (12) | (13) |
| 81 | NEP175636+663911 | 17:56:36.5 | +66:39:11.8 | 0.18 | 0.25 | 0.152 | 0.036 | 0.318 | 0.038 | 25.2 | 13.9 | 31.4 |
| 82 | NEP175636+663117 | 17:56:36.5 | +66:31:17.4 | 2.16 | 0.98 | 0.109 | 0.027 | 0.344 | 0.031 | 30.5 | 22.9 | 70.5 |
| 83 | NEP175636+671349 | 17:56:36.7 | +67:13:49.6 | 0.04 | 0.04 | 4.100 | 0.417 | 6.276 | 0.442 | 17.9 | 10.4 | -36.2 |
| 84 | NEP175637+664025 | 17:56:37.1 | +66:40:25.9 | 0.04 | 0.04 | 0.396 | 0.036 | 0.441 | 0.039 | | | |
| 85 | NEP175638+663147 | 17:56:38.8 | +66:31:47.1 | 0.06 | 0.03 | 0.133 | 0.027 | 0.119 | 0.027 | | | |
| 86 | NEP175639+664800 | 17:56:39.9 | +66:48:00.8 | 0.00 | 0.00 | 0.672 | 0.038 | 0.706 | 0.039 | | | |
| 87 | NEP175640+670444 | 17:56:40.6 | +67:04:44.3 | 0.00 | 0.00 | 8.093 | 0.100 | 10.113 | 0.138 | | | |
| 88 | NEP175640+663416 | 17:56:40.9 | +66:34:16.6 | 0.00 | 0.00 | 0.296 | 0.034 | 0.250 | 0.035 | | | |
| 89 | NEP175641+661528 | 17:56:41.4 | +66:15:28.1 | 0.00 | 0.00 | 1.681 | 0.040 | 1.701 | 0.044 | | | |
| 90 | NEP175642+663333 | 17:56:42.5 | +66:33:33.2 | 0.00 | 0.00 | 40.429 | 0.042 | 47.392 | 0.192 | | | |
| 91 | NEP175644+663738 | 17:56:44.2 | +66:37:38.1 | 0.06 | 0.03 | 0.197 | 0.033 | 0.164 | 0.034 | | | |
| 92 | NEP175648+662737 | 17:56:48.8 | +66:27:37.9 | 0.09 | 0.08 | 0.687 | 0.031 | 0.868 | 0.061 | | | |
| 93 | NEP175651+661308 | 17:56:51.4 | +66:13:08.2 | 0.03 | 0.03 | 0.324 | 0.057 | 0.351 | 0.058 | | | |
| 94 | NEP175651+662058 | 17:56:51.4 | +66:20:58.8 | 0.04 | 0.04 | 0.202 | 0.032 | 0.147 | 0.033 | | | |
| 95 | NEP175652+661144 | 17:56:52.6 | +66:11:44.5 | 0.02 | 0.01 | 1.348 | 0.053 | 2.346 | 0.066 | 21.9 | 10.5 | 59.5 |
| 96 | NEP175653+664532 | 17:56:53.9 | +66:45:33.0 | 0.02 | 0.02 | 0.862 | 0.033 | 0.949 | 0.047 | | | |
| 97 | NEP175655+662940 | 17:56:55.9 | +66:29:40.1 | 0.03 | 0.04 | 0.176 | 0.031 | 0.190 | 0.032 | | | |
| 98 | NEP175657+655745 | 17:56:57.1 | +65:57:45.0 | 0.04 | 0.05 | 0.596 | 0.113 | 0.723 | 0.115 | | | |
| 99 | NEP175658+662358 | 17:56:58.7 | +66:23:58.9 | 0.07 | 0.08 | 0.170 | 0.033 | 0.189 | 0.034 | | | |
| 100 | NEP175702+661328 | 17:57:02.3 | +66:13:29.0 | 0.00 | 0.00 | 7.180 | 0.057 | 7.652 | 0.067 | | | |

Table 1. —continued

| Running number | Source name | RA hh:mm:ss.s | DEC dd:mm:ss.s | δRA " | δDEC " | $S_{peak}$ mJy beam$^{-1}$ | $S_{peak}$ error mJy beam$^{-1}$ | $S_{total}$ mJy | $S_{total}$ error mJy | $\theta_{maj}$ " | $\theta_{min}$ " | PA ° |
|---|---|---|---|---|---|---|---|---|---|---|---|---|
| (1) | (2) | (3) | (4) | (5) | (6) | (7) | (8) | (9) | (10) | (11) | (12) | (13) |
| 101 | NEP175702+662404 | 17:57:02.4 | +66:24:04.4 | 0.02 | 0.03 | 0.208 | 0.033 | 0.178 | 0.034 | | | |
| 102 | NEP175703+670910 | 17:57:03.3 | +67:09:10.4 | 0.00 | 0.00 | 2.493 | 0.190 | 2.746 | 0.199 | | | |
| 103 | NEP175703+663058 | 17:57:03.5 | +66:30:58.0 | 0.01 | 0.02 | 0.191 | 0.034 | 0.148 | 0.034 | | | |
| 104 | NEP175703+665418 | 17:57:03.8 | +66:54:18.8 | 0.00 | 0.00 | 27.315 | 0.069 | 32.128 | 0.113 | | | |
| 105 | NEP175706+664903 | 17:57:06.4 | +66:49:03.4 | 0.03 | 0.05 | 0.253 | 0.038 | 0.217 | 0.040 | | | |
| 106 | NEP175707+664132 | 17:57:07.7 | +66:41:32.7 | 0.07 | 0.23 | 0.230 | 0.031 | 0.458 | 0.036 | 29.6 | 5.9 | 3.8 |
| 107 | NEP175707+661917 | 17:57:07.8 | +66:19:17.5 | 0.01 | 0.02 | 1.130 | 0.032 | 1.523 | 0.049 | | | |
| 108 | NEP175711+662221 | 17:57:11.7 | +66:22:21.1 | 0.01 | 0.02 | 0.359 | 0.032 | 0.328 | 0.033 | | | |
| 109 | NEP175714+662139 | 17:57:14.4 | +66:21:39.2 | 0.00 | 0.00 | 0.156 | 0.027 | 0.096 | 0.027 | | | |
| 110 | NEP175714+662529 | 17:57:14.5 | +66:25:29.0 | 0.01 | 0.01 | 0.184 | 0.030 | 0.129 | 0.030 | | | |
| 111 | NEP175715+665449 | 17:57:15.4 | +66:54:49.6 | 0.01 | 0.04 | 0.629 | 0.046 | 0.641 | 0.051 | | | |
| 112 | NEP175715+661605 | 17:57:15.7 | +66:16:05.2 | 0.27 | 0.17 | 0.210 | 0.037 | 0.611 | 0.041 | 31.0 | 20.1 | -61.6 |
| 113 | NEP175718+664726 | 17:57:18.9 | +66:47:26.9 | 0.15 | 0.26 | 0.179 | 0.038 | 0.244 | 0.041 | 18.6 | 2.2 | -9.6 |
| 114 | NEP175721+662945 | 17:57:21.2 | +66:29:45.8 | 0.01 | 0.01 | 0.467 | 0.031 | 0.421 | 0.032 | | | |
| 115 | NEP175722+670332 | 17:57:22.2 | +67:03:32.1 | 0.02 | 0.05 | 0.627 | 0.092 | 0.808 | 0.096 | 19.7 | | -13.5 |
| 116 | NEP175725+661546 | 17:57:25.7 | +66:15:46.0 | 0.03 | 0.04 | 0.295 | 0.037 | 0.364 | 0.038 | | | |
| 117 | NEP175727+664557 | 17:57:27.7 | +66:45:57.1 | 0.01 | 0.01 | 0.281 | 0.035 | 0.229 | 0.035 | | | |
| 118 | NEP175728+663822 | 17:57:28.2 | +66:38:22.1 | 0.10 | 0.06 | 0.674 | 0.030 | 1.077 | 0.055 | 18.8 | 9.8 | 65.4 |
| 119 | NEP175729+660947 | 17:57:29.0 | +66:09:47.9 | 0.09 | 0.19 | 0.293 | 0.051 | 0.387 | 0.055 | 19.9 | | 16.4 |
| 120 | NEP175729+662417 | 17:57:29.4 | +66:24:17.1 | 0.03 | 0.03 | 0.215 | 0.030 | 0.175 | 0.031 | | | |

Table 1. *−continued*

| Running number | Source name | RA hh:mm:ss.s | DEC dd:mm:ss.s | δRA " | δDEC " | $S_{peak}$ mJy beam$^{-1}$ | $S_{peak}$ error mJy beam$^{-1}$ | $S_{total}$ mJy | $S_{total}$ error mJy | $\theta_{maj}$ " | $\theta_{min}$ " | PA ° |
|---|---|---|---|---|---|---|---|---|---|---|---|---|
| (1) | (2) | (3) | (4) | (5) | (6) | (7) | (8) | (9) | (10) | (11) | (12) | (13) |
| 121 | NEP175731+670433 | 17:57:31.7 | +67:04:33.7 | 0.00 | 0.00 | 17.560 | 0.092 | 21.016 | 0.176 | | | |
| 122 | NEP175732+663625 | 17:57:32.3 | +66:36:25.8 | 0.91 | 0.18 | 0.129 | 0.030 | 0.353 | 0.033 | 39.8 | 12.1 | 89.3 |
| 123 | NEP175732+654853 | 17:57:33.0 | +65:48:53.7 | 0.02 | 0.32 | 1.953 | 0.372 | 2.477 | 0.389 | 19.5 | | 13.7 |
| 124 | NEP175734+654859 | 17:57:34.4 | +65:48:59.3 | 1.33 | 1.90 | 1.603 | 0.372 | 4.732 | 0.412 | 30.8 | 20.6 | 57.2 |
| 125 | NEP175735+662206 | 17:57:35.3 | +66:22:06.8 | 0.01 | 0.02 | 0.174 | 0.027 | 0.139 | 0.027 | | | |
| 126 | NEP175737+665439 | 17:57:37.5 | +66:54:39.6 | 0.00 | 0.00 | 0.709 | 0.046 | 0.733 | 0.048 | | | |
| 127 | NEP175737+664443 | 17:57:37.8 | +66:44:43.3 | 0.01 | 0.01 | 4.166 | 0.035 | 5.486 | 0.130 | | | |
| 128 | NEP175740+662830 | 17:57:40.0 | +66:28:30.6 | 0.01 | 0.02 | 0.809 | 0.031 | 0.813 | 0.040 | | | |
| 129 | NEP175740+664550 | 17:57:40.3 | +66:45:50.3 | 0.05 | 0.22 | 1.364 | 0.036 | 2.976 | 0.113 | 31.0 | 6.1 | 12.5 |
| 130 | NEP175743+662644 | 17:57:43.8 | +66:26:44.4 | 0.02 | 0.02 | 0.177 | 0.027 | 0.125 | 0.027 | | | |
| 131 | NEP175744+663718 | 17:57:44.1 | +66:37:18.4 | 0.18 | 0.15 | 0.138 | 0.030 | 0.309 | 0.031 | 28.9 | 13.4 | -43.5 |
| 132 | NEP175744+664358 | 17:57:44.4 | +66:43:58.3 | 0.04 | 0.03 | 0.197 | 0.036 | 0.132 | 0.037 | | | |
| 133 | NEP175744+660932 | 17:57:44.8 | +66:09:32.8 | 0.00 | 0.01 | 1.916 | 0.054 | 2.085 | 0.064 | | | |
| 134 | NEP175745+662904 | 17:57:45.4 | +66:29:04.8 | 0.02 | 0.04 | 0.169 | 0.025 | 0.126 | 0.025 | | | |
| 135 | NEP175746+665437 | 17:57:46.5 | +66:54:37.2 | 0.02 | 0.02 | 0.275 | 0.044 | 0.452 | 0.045 | 21.8 | 9.3 | -39.1 |
| 136 | NEP175746+661022 | 17:57:47.0 | +66:10:22.9 | 0.04 | 0.07 | 0.325 | 0.054 | 0.366 | 0.056 | | | |
| 137 | NEP175748+665903 | 17:57:48.5 | +66:59:03.7 | 0.02 | 0.08 | 0.622 | 0.045 | 0.932 | 0.052 | 25.8 | | 8.8 |
| 138 | NEP175748+655340 | 17:57:48.9 | +65:53:40.7 | 0.01 | 0.02 | 2.741 | 0.214 | 3.354 | 0.227 | | | |
| 139 | NEP175749+663132 | 17:57:49.7 | +66:31:32.8 | 0.04 | 0.04 | 0.158 | 0.027 | 0.289 | 0.028 | 20.0 | 14.9 | 7.0 |
| 140 | NEP175753+670042 | 17:57:53.6 | +67:00:42.2 | 0.00 | 0.00 | 0.551 | 0.065 | 0.449 | 0.065 | | | |

Table 1. — continued

| Running number | Source name | RA hh:mm:ss.s | DEC dd:mm:ss.s | δRA " | δDEC " | $S_{peak}$ mJy beam$^{-1}$ | $S_{peak}$ error mJy beam$^{-1}$ | $S_{total}$ mJy | $S_{total}$ error mJy | $\theta_{maj}$ " | $\theta_{min}$ " | PA ° |
|---|---|---|---|---|---|---|---|---|---|---|---|---|
| (1) | (2) | (3) | (4) | (5) | (6) | (7) | (8) | (9) | (10) | (11) | (12) | (13) |
| 141 | NEP175754+664857 | 17:57:54.9 | +66:48:57.1 | 0.00 | 0.00 | 3.588 | 0.040 | 3.792 | 0.050 | | | |
| 142 | NEP175755+662857 | 17:57:55.5 | +66:28:57.3 | 0.02 | 0.03 | 0.141 | 0.025 | 0.124 | 0.025 | | | |
| 143 | NEP175756+660925 | 17:57:56.7 | +66:09:25.8 | 0.01 | 0.01 | 0.822 | 0.054 | 0.835 | 0.057 | | | |
| 144 | NEP175758+662927 | 17:57:59.0 | +66:29:27.4 | 0.01 | 0.01 | 0.206 | 0.025 | 0.155 | 0.025 | | | |
| 145 | NEP175803+663547 | 17:58:03.0 | +66:35:47.6 | 0.09 | 0.11 | 0.455 | 0.030 | 0.683 | 0.043 | 18.3 | 7.6 | 32.3 |
| 146 | NEP175804+661103 | 17:58:04.5 | +66:11:03.9 | 0.02 | 0.02 | 0.281 | 0.054 | 0.233 | 0.054 | | | |
| 147 | NEP175805+663620 | 17:58:05.4 | +66:36:20.3 | 0.50 | 0.16 | 0.186 | 0.030 | 0.436 | 0.036 | 31.0 | 12.8 | -72.8 |
| 148 | NEP175805+662527 | 17:58:05.5 | +66:25:27.2 | 0.01 | 0.01 | 0.177 | 0.027 | 0.123 | 0.027 | | | |
| 149 | NEP175807+663640 | 17:58:07.9 | +66:36:40.4 | 1.42 | 0.26 | 0.149 | 0.030 | 0.317 | 0.034 | 27.5 | 13.0 | 82.6 |
| 150 | NEP175808+664909 | 17:58:08.4 | +66:49:09.9 | 0.00 | 0.00 | 13.260 | 0.040 | 13.573 | 0.058 | | | |
| 151 | NEP175809+661532 | 17:58:09.4 | +66:15:32.1 | 0.00 | 0.00 | 1.050 | 0.032 | 1.073 | 0.037 | | | |
| 152 | NEP175809+664134 | 17:58:09.4 | +66:41:34.9 | 0.00 | 0.00 | 7.367 | 0.036 | 10.512 | 0.094 | 18.1 | 2.7 | -21.9 |
| 153 | NEP175813+661516 | 17:58:13.3 | +66:15:16.1 | 0.02 | 0.01 | 0.179 | 0.028 | 0.142 | 0.028 | | | |
| 154 | NEP175813+661225 | 17:58:13.8 | +66:12:25.6 | 0.03 | 0.05 | 0.242 | 0.047 | 0.280 | 0.048 | | | |
| 155 | NEP175814+664425 | 17:58:14.4 | +66:44:25.2 | 0.00 | 0.00 | 3.196 | 0.031 | 3.185 | 0.040 | | | |
| 156 | NEP175816+661149 | 17:58:16.3 | +66:11:49.3 | 0.04 | 0.02 | 1.088 | 0.048 | 1.963 | 0.068 | 20.8 | 12.6 | 70.7 |
| 157 | NEP175816+661006 | 17:58:16.8 | +66:10:06.1 | 0.00 | 0.00 | 0.563 | 0.048 | 0.479 | 0.048 | | | |
| 158 | NEP175818+664325 | 17:58:18.3 | +66:43:25.8 | 0.03 | 0.02 | 0.167 | 0.031 | 0.144 | 0.031 | | | |
| 159 | NEP175819+661231 | 17:58:19.7 | +66:12:31.1 | 0.00 | 0.00 | 3.046 | 0.047 | 3.451 | 0.051 | | | |
| 160 | NEP175820+670207 | 17:58:20.2 | +67:02:07.3 | 0.07 | 0.14 | 0.413 | 0.080 | 0.444 | 0.084 | | | |

Table 1. – *continued*

| Running number | Source name | RA hh:mm:ss.s | DEC dd:mm:ss.s | δRA " | δDEC " | $S_{peak}$ mJy beam$^{-1}$ | $S_{peak}$ error mJy beam$^{-1}$ | $S_{total}$ mJy | $S_{total}$ error mJy | $\theta_{maj}$ " | $\theta_{min}$ " | PA ° |
|---|---|---|---|---|---|---|---|---|---|---|---|---|
| (1) | (2) | (3) | (4) | (5) | (6) | (7) | (8) | (9) | (10) | (11) | (12) | (13) |
| 161 | NEP175821+662855 | 17:58:21.3 | +66:28:55.9 | 0.00 | 0.00 | 1.010 | 0.026 | 1.087 | 0.028 | | | |
| 162 | NEP175821+661349 | 17:58:21.6 | +66:13:49.5 | 0.00 | 0.00 | 6.011 | 0.047 | 6.883 | 0.079 | | | |
| 163 | NEP175823+662121 | 17:58:23.6 | +66:21:21.6 | 0.01 | 0.02 | 0.167 | 0.029 | 0.151 | 0.029 | | | |
| 164 | NEP175824+663912 | 17:58:24.1 | +66:39:12.7 | 0.58 | 0.40 | 0.212 | 0.036 | 0.293 | 0.041 | 23.8 | | 52.9 |
| 165 | NEP175824+665814 | 17:58:24.6 | +66:58:14.9 | 0.04 | 0.07 | 0.615 | 0.056 | 0.604 | 0.061 | | | |
| 166 | NEP175824+664420 | 17:58:24.8 | +66:44:20.3 | 0.00 | 0.01 | 0.385 | 0.031 | 0.283 | 0.032 | | | |
| 167 | NEP175826+670148 | 17:58:26.9 | +67:01:48.5 | 0.01 | 0.01 | 6.004 | 0.059 | 8.601 | 0.157 | | | |
| 168 | NEP175826+655505 | 17:58:27.0 | +65:55:05.9 | 0.10 | 0.07 | 21.968 | 3.026 | 191.847 | 3.257 | 52.5 | 48.2 | 82.5 |
| 169 | NEP175827+665112 | 17:58:27.3 | +66:51:12.8 | 0.03 | 0.04 | 0.293 | 0.042 | 0.309 | 0.043 | | | |
| 170 | NEP175827+665537 | 17:58:27.5 | +66:55:37.5 | 0.02 | 0.03 | 0.270 | 0.040 | 0.237 | 0.041 | | | |
| 171 | NEP175828+665811 | 17:58:28.7 | +66:58:11.6 | 1.06 | 0.42 | 0.210 | 0.056 | 0.672 | 0.062 | 30.4 | 24.0 | 87.0 |
| 172 | NEP175830+664926 | 17:58:30.1 | +66:49:26.2 | 0.00 | 0.01 | 1.668 | 0.045 | 2.864 | 0.055 | 24.3 | 5.8 | 23.5 |
| 173 | NEP175830+663932 | 17:58:31.0 | +66:39:32.6 | 0.25 | 0.14 | 0.197 | 0.036 | 0.264 | 0.041 | | | |
| 174 | NEP175831+661145 | 17:58:31.1 | +66:11:45.1 | 0.01 | 0.03 | 0.304 | 0.048 | 0.245 | 0.049 | | | |
| 175 | NEP175831+662318 | 17:58:31.9 | +66:23:18.4 | 0.00 | 0.00 | 1.237 | 0.029 | 1.326 | 0.030 | | | |
| 176 | NEP175832+662722 | 17:58:32.8 | +66:27:23.0 | 0.25 | 0.28 | 0.130 | 0.026 | 0.220 | 0.027 | 29.2 | | -48.9 |
| 177 | NEP175832+655259 | 17:58:33.0 | +65:52:59.6 | 0.01 | 0.01 | 12.632 | 0.742 | 76.257 | 0.829 | 43.3 | 37.5 | 40.4 |
| 178 | NEP175833+663759 | 17:58:33.3 | +66:37:59.5 | 0.00 | 0.00 | 509.459 | 1.133 | 786.233 | 2.568 | | | |
| 179 | NEP175835+662953 | 17:58:35.3 | +66:29:53.2 | 0.03 | 0.02 | 0.140 | 0.029 | 0.128 | 0.029 | | | |
| 180 | NEP175835+663843 | 17:58:36.0 | +66:38:43.9 | 0.51 | 0.93 | 0.208 | 0.036 | 0.393 | 0.038 | 27.6 | 8.6 | -49.4 |

Table 1. – *continued*

| Running number | Source name | RA hh:mm:ss.s | DEC dd:mm:ss.s | δRA " | δDEC " | $S_{peak}$ mJy beam$^{-1}$ | $S_{peak}$ error mJy beam$^{-1}$ | $S_{total}$ mJy | $S_{total}$ error mJy | $\theta_{maj}$ " | $\theta_{min}$ " | PA ° |
|---|---|---|---|---|---|---|---|---|---|---|---|---|
| (1) | (2) | (3) | (4) | (5) | (6) | (7) | (8) | (9) | (10) | (11) | (12) | (13) |
| 181 | NEP175838+663733 | 17:58:38.6 | +66:37:33.4 | **** | 6.24 | 0.545 | 1.133 | 1.881 | 1.134 | 36.8 | 24.4 | -56.5 |
| 182 | NEP175840+663727 | 17:58:40.3 | +66:37:27.6 | 4.30 | 1.09 | 0.317 | 0.044 | 1.045 | 0.057 | 33.8 | 20.9 | 18.8 |
| 183 | NEP175840+663920 | 17:58:40.6 | +66:39:20.9 | 0.16 | 0.24 | 0.156 | 0.033 | 0.273 | 0.035 | 26.6 | 5.2 | -28.5 |
| 184 | NEP175840+662212 | 17:58:40.7 | +66:22:12.5 | 0.08 | 0.08 | 0.350 | 0.030 | 0.514 | 0.036 | | | |
| 185 | NEP175843+663419 | 17:58:43.6 | +66:34:19.8 | 0.00 | 0.01 | 0.189 | 0.028 | 0.123 | 0.028 | | | |
| 186 | NEP175846+661341 | 17:58:46.3 | +66:13:41.6 | 0.00 | 0.00 | 0.409 | 0.035 | 0.387 | 0.035 | | | |
| 187 | NEP175846+660732 | 17:58:46.4 | +66:07:32.2 | 0.04 | 0.03 | 0.261 | 0.045 | 0.210 | 0.046 | | | |
| 188 | NEP175846+664759 | 17:58:46.6 | +66:47:59.7 | 0.01 | 0.01 | 0.262 | 0.035 | 0.208 | 0.035 | | | |
| 189 | NEP175847+661418 | 17:58:47.9 | +66:14:18.9 | 0.01 | 0.01 | 0.455 | 0.035 | 0.466 | 0.036 | | | |
| 190 | NEP175848+660949 | 17:58:48.9 | +66:09:49.2 | 0.02 | 0.04 | 0.357 | 0.040 | 0.390 | 0.041 | | | |
| 191 | NEP175849+660716 | 17:58:49.7 | +66:07:16.2 | 0.01 | 0.01 | 0.634 | 0.045 | 0.647 | 0.047 | | | |
| 192 | NEP175850+663736 | 17:58:50.1 | +66:37:36.8 | 0.10 | 0.12 | 0.246 | 0.044 | 0.264 | 0.047 | | | |
| 193 | NEP175852+662934 | 17:58:52.1 | +66:29:34.8 | 0.00 | 0.00 | 3.549 | 0.026 | 3.517 | 0.035 | | | |
| 194 | NEP175852+662008 | 17:58:52.6 | +66:20:08.7 | 0.00 | 0.00 | 1.665 | 0.031 | 1.818 | 0.043 | | | |
| 195 | NEP175853+670636 | 17:58:53.2 | +67:06:36.0 | 0.01 | 0.01 | 5.382 | 0.105 | 7.120 | 0.163 | | | |
| 196 | NEP175853+662320 | 17:58:53.3 | +66:23:20.9 | 0.09 | 0.05 | 0.148 | 0.030 | 0.174 | 0.031 | | | |
| 197 | NEP175856+661912 | 17:58:56.6 | +66:19:12.8 | 0.12 | 0.17 | 0.156 | 0.031 | 0.268 | 0.033 | 19.2 | 13.0 | -1.0 |
| 198 | NEP175857+664932 | 17:58:57.3 | +66:49:32.2 | 0.00 | 0.01 | 1.651 | 0.035 | 1.877 | 0.047 | | | |
| 199 | NEP175859+663017 | 17:58:59.6 | +66:30:17.5 | 0.02 | 0.02 | 0.435 | 0.028 | 0.419 | 0.031 | | | |
| 200 | NEP175904+661339 | 17:59:04.9 | +66:13:39.5 | 0.00 | 0.00 | 1.745 | 0.035 | 1.715 | 0.041 | | | |

Table 1. − *continued*

| Running number | Source name | RA hh:mm:ss.s | DEC dd:mm:ss.s | δRA " | δDEC " | $S_{peak}$ mJy beam$^{-1}$ | $S_{peak}$ error mJy beam$^{-1}$ | $S_{total}$ mJy | $S_{total}$ error mJy | $\theta_{maj}$ " | $\theta_{min}$ " | PA ° |
|---|---|---|---|---|---|---|---|---|---|---|---|---|
| (1) | (2) | (3) | (4) | (5) | (6) | (7) | (8) | (9) | (10) | (11) | (12) | (13) |
| 201 | NEP175905+662144 | 17:59:05.1 | +66:21:45.0 | 0.01 | 0.01 | 0.738 | 0.030 | 0.754 | 0.034 | | | |
| 202 | NEP175905+661605 | 17:59:05.3 | +66:16:05.8 | 0.00 | 0.01 | 1.787 | 0.031 | 1.870 | 0.048 | | | |
| 203 | NEP175905+664804 | 17:59:05.8 | +66:48:04.8 | 0.00 | 0.01 | 0.342 | 0.035 | 0.312 | 0.035 | | | |
| 204 | NEP175905+663229 | 17:59:05.9 | +66:32:29.8 | 0.00 | 0.00 | 0.309 | 0.028 | 0.281 | 0.028 | | | |
| 205 | NEP175908+662516 | 17:59:08.3 | +66:25:16.1 | 0.00 | 0.00 | 0.474 | 0.027 | 0.450 | 0.027 | | | |
| 206 | NEP175909+663924 | 17:59:09.8 | +66:39:24.5 | 0.00 | 0.00 | 1.038 | 0.035 | 0.995 | 0.036 | | | |
| 207 | NEP175910+664002 | 17:59:10.4 | +66:40:02.9 | 0.01 | 0.01 | 0.458 | 0.035 | 0.481 | 0.036 | | | |
| 208 | NEP175911+660427 | 17:59:11.1 | +66:04:27.6 | 0.01 | 0.01 | 0.482 | 0.062 | 0.517 | 0.063 | | | |
| 209 | NEP175911+665014 | 17:59:11.1 | +66:50:14.3 | 0.00 | 0.00 | 1.021 | 0.035 | 0.987 | 0.038 | | | |
| 210 | NEP175911+663004 | 17:59:11.1 | +66:30:04.9 | 0.01 | 0.02 | 0.256 | 0.029 | 0.235 | 0.029 | | | |
| 211 | NEP175911+660450 | 17:59:11.6 | +66:04:51.0 | 0.42 | 1.24 | 0.327 | 0.062 | 0.555 | 0.069 | 23.0 | 8.0 | -30.8 |
| 212 | NEP175911+663348 | 17:59:11.9 | +66:33:49.0 | 0.00 | 0.00 | 1.091 | 0.028 | 0.975 | 0.032 | | | |
| 213 | NEP175915+663302 | 17:59:15.4 | +66:33:02.8 | 0.01 | 0.01 | 0.769 | 0.028 | 0.784 | 0.034 | | | |
| 214 | NEP175916+661031 | 17:59:16.8 | +66:10:31.7 | 0.02 | 0.02 | 0.348 | 0.043 | 0.344 | 0.044 | | | |
| 215 | NEP175919+660636 | 17:59:19.7 | +66:06:36.3 | 0.03 | 0.04 | 0.464 | 0.053 | 0.609 | 0.055 | | | |
| 216 | NEP175921+664235 | 17:59:21.4 | +66:42:35.1 | 0.01 | 0.01 | 0.259 | 0.032 | 0.188 | 0.032 | | | |
| 217 | NEP175921+663858 | 17:59:21.6 | +66:38:58.6 | 0.01 | 0.01 | 0.959 | 0.035 | 0.931 | 0.041 | | | |
| 218 | NEP175922+665036 | 17:59:22.4 | +66:50:36.3 | 0.00 | 0.00 | 0.956 | 0.036 | 1.009 | 0.039 | | | |
| 219 | NEP175923+654958 | 17:59:23.1 | +65:49:58.7 | 0.03 | 0.06 | 3.310 | 0.278 | 4.754 | 0.312 | 19.5 | 3.8 | 5.2 |
| 220 | NEP175926+661440 | 17:59:26.0 | +66:14:40.4 | 0.16 | 0.08 | 0.151 | 0.030 | 0.426 | 0.031 | 30.4 | 20.3 | 77.1 |

Table 1. — continued

| Running number | Source name | RA hh:mm:ss.s | DEC dd:mm:ss.s | δRA " | δDEC " | $S_{peak}$ mJy beam$^{-1}$ | $S_{peak}$ error mJy beam$^{-1}$ | $S_{total}$ mJy | $S_{total}$ error mJy | $\theta_{maj}$ " | $\theta_{min}$ " | PA ° |
|---|---|---|---|---|---|---|---|---|---|---|---|---|
| (1) | (2) | (3) | (4) | (5) | (6) | (7) | (8) | (9) | (10) | (11) | (12) | (13) |
| 221 | NEP175926+662941 | 17:59:26.4 | +66:29:41.0 | 0.01 | 0.04 | 0.170 | 0.029 | 0.168 | 0.029 | | | |
| 222 | NEP175927+664229 | 17:59:27.6 | +66:42:29.9 | 0.01 | 0.02 | 0.275 | 0.032 | 0.276 | 0.033 | | | |
| 223 | NEP175929+663227 | 17:59:29.1 | +66:32:27.2 | 0.01 | 0.01 | 0.222 | 0.029 | 0.210 | 0.029 | | | |
| 224 | NEP175930+655642 | 17:59:30.6 | +65:56:42.9 | 0.00 | 0.01 | 2.865 | 0.139 | 3.325 | 0.148 | | | |
| 225 | NEP175931+662906 | 17:59:31.8 | +66:29:06.9 | 0.00 | 0.00 | 0.269 | 0.029 | 0.247 | 0.029 | | | |
| 226 | NEP175933+661226 | 17:59:33.5 | +66:12:26.5 | 0.43 | 0.15 | 0.142 | 0.030 | 0.765 | 0.033 | 51.1 | 27.5 | 89.9 |
| 227 | NEP175934+662704 | 17:59:34.4 | +66:27:04.5 | 0.00 | 0.00 | 0.430 | 0.029 | 0.381 | 0.030 | | | |
| 228 | NEP175934+670214 | 17:59:34.5 | +67:02:14.2 | 0.00 | 0.00 | 2.351 | 0.074 | 2.678 | 0.083 | | | |
| 229 | NEP175934+661306 | 17:59:34.6 | +66:13:06.5 | 0.01 | 0.01 | 0.361 | 0.030 | 0.342 | 0.031 | | | |
| 230 | NEP175935+662104 | 17:59:35.5 | +66:21:04.3 | 0.61 | 1.12 | 0.117 | 0.028 | 0.270 | 0.030 | 26.9 | 15.4 | -12.5 |
| 231 | NEP175936+661604 | 17:59:36.2 | +66:16:04.4 | 0.00 | 0.00 | 20.371 | 0.132 | 21.707 | 0.207 | | | |
| 232 | NEP175936+671856 | 17:59:36.7 | +67:18:56.3 | 0.03 | 0.14 | 2.504 | 0.416 | 2.346 | 0.421 | | | |
| 233 | NEP175937+661316 | 17:59:37.8 | +66:13:16.8 | 1.16 | 2.04 | 0.123 | 0.030 | 0.659 | 0.034 | 40.6 | 34.8 | 28.0 |
| 234 | NEP175937+671726 | 17:59:37.8 | +67:17:26.6 | 0.01 | 0.01 | 9.394 | 0.475 | 14.724 | 0.518 | 20.4 | 7.6 | 3.7 |
| 235 | NEP175938+655841 | 17:59:38.1 | +65:58:41.7 | 0.00 | 0.00 | 1.054 | 0.094 | 0.368 | 0.095 | | | |
| 236 | NEP175940+665155 | 17:59:40.6 | +66:51:55.7 | 0.02 | 0.03 | 0.200 | 0.037 | 0.197 | 0.037 | | | |
| 237 | NEP175940+664028 | 17:59:40.7 | +66:40:28.2 | 0.00 | 0.00 | 0.191 | 0.024 | 0.145 | 0.024 | | | |
| 238 | NEP175942+662107 | 17:59:42.4 | +66:21:08.0 | 0.12 | 0.09 | 0.150 | 0.030 | 0.165 | 0.032 | | | |
| 239 | NEP175942+662722 | 17:59:42.7 | +66:27:22.6 | 0.04 | 0.04 | 0.178 | 0.027 | 0.181 | 0.028 | | | |
| 240 | NEP175943+662452 | 17:59:43.1 | +66:24:52.7 | 0.11 | 0.15 | 0.133 | 0.022 | 0.126 | 0.024 | | | |

Table 1. −continued

| Running number | Source name | RA hh:mm:ss.s | DEC dd:mm:ss.s | δRA " | δDEC " | $S_{peak}$ mJy beam$^{-1}$ | $S_{peak}$ error mJy beam$^{-1}$ | $S_{total}$ mJy | $S_{total}$ error mJy | $\theta_{maj}$ " | $\theta_{min}$ " | PA ° |
|---|---|---|---|---|---|---|---|---|---|---|---|---|
| (1) | (2) | (3) | (4) | (5) | (6) | (7) | (8) | (9) | (10) | (11) | (12) | (13) |
| 241 | NEP175948+661457 | 17:59:48.3 | +66:14:57.1 | 0.07 | 0.06 | 0.174 | 0.034 | 0.189 | 0.035 | | | |
| 242 | NEP175953+664705 | 17:59:53.3 | +66:47:05.3 | 0.02 | 0.02 | 0.828 | 0.030 | 0.880 | 0.042 | | | |
| 243 | NEP175954+670218 | 17:59:55.0 | +67:02:18.1 | 0.05 | 0.10 | 0.370 | 0.078 | 0.533 | 0.080 | 21.9 | | 23.8 |
| 244 | NEP175958+664647 | 17:59:59.0 | +66:46:47.9 | 0.15 | 0.24 | 0.181 | 0.030 | 0.285 | 0.034 | 18.0 | 10.7 | 16.9 |
| 245 | NEP175959+670249 | 17:59:59.1 | +67:02:49.3 | 0.01 | 0.01 | 1.664 | 0.078 | 2.374 | 0.087 | 18.0 | 7.1 | 40.5 |
| 246 | NEP18000+663110 | 18:00:0.2 | +66:31:10.2 | 0.03 | 0.02 | 0.165 | 0.030 | 0.180 | 0.030 | | | |
| 247 | NEP180002+661045 | 18:00:02.4 | +66:10:45.8 | 0.04 | 0.02 | 0.269 | 0.042 | 0.243 | 0.043 | | | |
| 248 | NEP180004+662257 | 18:00:04.8 | +66:22:58.0 | 0.02 | 0.02 | 0.411 | 0.030 | 0.481 | 0.033 | | | |
| 249 | NEP180005+661040 | 18:00:05.7 | +66:10:41.0 | 0.03 | 0.41 | 0.385 | 0.042 | 0.111 | 0.042 | | | |
| 250 | NEP180007+663654 | 18:00:07.1 | +66:36:54.9 | 0.00 | 0.00 | 29.952 | 0.032 | 31.675 | 0.078 | | | |
| 251 | NEP180008+664500 | 18:00:08.9 | +66:45:00.6 | 0.33 | 0.30 | 0.148 | 0.034 | 0.364 | 0.037 | 26.0 | 18.6 | 49.3 |
| 252 | NEP180010+661943 | 18:00:10.5 | +66:19:43.5 | 0.01 | 0.01 | 3.339 | 0.032 | 5.475 | 0.103 | 17.2 | 11.6 | -72.3 |
| 253 | NEP180011+663326 | 18:00:11.3 | +66:33:26.3 | 0.02 | 0.04 | 0.146 | 0.028 | 0.154 | 0.028 | | | |
| 254 | NEP180011+665213 | 18:00:11.4 | +66:52:13.6 | 0.15 | 0.33 | 0.450 | 0.032 | 0.743 | 0.051 | 22.6 | 5.0 | 16.3 |
| 255 | NEP180016+662033 | 18:00:16.6 | +66:20:33.3 | 0.06 | 0.05 | 0.174 | 0.032 | 0.218 | 0.033 | | | |
| 256 | NEP180019+662404 | 18:00:19.6 | +66:24:04.7 | 0.18 | 0.31 | 0.109 | 0.023 | 0.209 | 0.025 | 25.5 | 10.1 | 18.7 |
| 257 | NEP180020+662226 | 18:00:20.3 | +66:22:26.8 | 0.00 | 0.00 | 1.317 | 0.031 | 1.302 | 0.037 | | | |
| 258 | NEP180021+670309 | 18:00:21.0 | +67:03:09.9 | 0.00 | 0.00 | 0.795 | 0.076 | 0.789 | 0.076 | | | |
| 259 | NEP180021+664212 | 18:00:21.7 | +66:42:12.9 | 0.02 | 0.03 | 0.161 | 0.029 | 0.124 | 0.029 | | | |
| 260 | NEP180022+663317 | 18:00:22.1 | +66:33:17.1 | 0.00 | 0.00 | 0.235 | 0.028 | 0.173 | 0.028 | | | |

Table 1. – *continued*

| Running number | Source name | RA hh:mm:ss.s | DEC dd:mm:ss.s | δRA " | δDEC " | $S_{peak}$ mJy beam$^{-1}$ | $S_{peak}$ error mJy beam$^{-1}$ | $S_{total}$ mJy | $S_{total}$ error mJy | $\theta_{maj}$ " | $\theta_{min}$ " | PA ° |
|---|---|---|---|---|---|---|---|---|---|---|---|---|
| (1) | (2) | (3) | (4) | (5) | (6) | (7) | (8) | (9) | (10) | (11) | (12) | (13) |
| 261 | NEP180023+660515 | 18:00:23.3 | +66:05:15.0 | 0.05 | 0.07 | 0.690 | 0.051 | 0.971 | 0.062 | | | |
| 262 | NEP180023+661550 | 18:00:23.8 | +66:15:50.0 | 0.05 | 0.32 | 0.172 | 0.031 | 0.293 | 0.034 | 31.0 | | 0.3 |
| 263 | NEP180027+665411 | 18:00:27.8 | +66:54:11.3 | 0.00 | 0.00 | 4.833 | 0.049 | 5.551 | 0.054 | | | |
| 264 | NEP180028+670552 | 18:00:28.7 | +67:05:52.9 | 0.02 | 0.03 | 4.347 | 0.089 | 5.736 | 0.205 | | | |
| 265 | NEP180028+664029 | 18:00:28.8 | +66:40:29.3 | 0.00 | 0.00 | 2.586 | 0.029 | 2.511 | 0.034 | | | |
| 266 | NEP180029+662920 | 18:00:29.4 | +66:29:20.3 | 0.07 | 0.15 | 0.118 | 0.025 | 0.123 | 0.026 | | | |
| 267 | NEP180030+670625 | 18:00:30.4 | +67:06:25.0 | 0.04 | 0.22 | 0.458 | 0.089 | 0.610 | 0.091 | 22.4 | | 3.3 |
| 268 | NEP180031+670453 | 18:00:31.2 | +67:04:53.5 | 0.01 | 0.02 | 0.557 | 0.076 | 0.620 | 0.077 | | | |
| 269 | NEP180031+664411 | 18:00:31.3 | +66:44:11.6 | 0.14 | 0.08 | 0.132 | 0.029 | 0.252 | 0.030 | 23.0 | 14.1 | -83.0 |
| 270 | NEP180032+661511 | 18:00:32.8 | +66:15:11.8 | 0.01 | 0.02 | 0.248 | 0.037 | 0.179 | 0.037 | | | |
| 271 | NEP180033+665248 | 18:00:33.0 | +66:52:48.7 | 0.01 | 0.02 | 0.164 | 0.032 | 0.179 | 0.032 | | | |
| 272 | NEP180033+664443 | 18:00:33.6 | +66:44:43.1 | 0.00 | 0.00 | 8.407 | 0.034 | 9.596 | 0.092 | | | |
| 273 | NEP180035+664941 | 18:00:35.6 | +66:49:41.7 | 0.01 | 0.01 | 0.217 | 0.032 | 0.210 | 0.033 | | | |
| 274 | NEP180035+662526 | 18:00:35.9 | +66:25:26.1 | 0.01 | 0.02 | 0.176 | 0.023 | 0.083 | 0.024 | | | |
| 275 | NEP180036+662207 | 18:00:36.4 | +66:22:07.8 | 0.52 | 0.17 | 0.140 | 0.031 | 0.332 | 0.034 | 30.6 | 14.0 | 79.7 |
| 276 | NEP180036+665221 | 18:00:36.6 | +66:52:21.8 | 0.01 | 0.01 | 0.705 | 0.032 | 0.640 | 0.035 | | | |
| 277 | NEP180037+664425 | 18:00:37.6 | +66:44:26.0 | 0.01 | 0.01 | 6.044 | 0.026 | 7.509 | 0.167 | | | |
| 278 | NEP180041+663355 | 18:00:41.7 | +66:33:55.6 | 0.09 | 0.05 | 0.405 | 0.031 | 0.477 | 0.037 | | | |
| 279 | NEP180042+664403 | 18:00:42.6 | +66:44:03.4 | 0.71 | 0.41 | 0.109 | 0.026 | 0.249 | 0.027 | 26.5 | 16.8 | 85.2 |
| 280 | NEP180047+661940 | 18:00:47.6 | +66:19:40.3 | 0.02 | 0.02 | 0.751 | 0.035 | 0.856 | 0.045 | | | |

Table 1. −continued

| Running number | Source name | RA hh:mm:ss.s | DEC dd:mm:ss.s | δRA " | δDEC " | $S_{peak}$ mJy beam$^{-1}$ | $S_{peak}$ error mJy beam$^{-1}$ | $S_{total}$ mJy | $S_{total}$ error mJy | $\theta_{maj}$ " | $\theta_{min}$ " | PA ° |
|---|---|---|---|---|---|---|---|---|---|---|---|---|
| (1) | (2) | (3) | (4) | (5) | (6) | (7) | (8) | (9) | (10) | (11) | (12) | (13) |
| 281 | NEP180050+662927 | 18:00:50.9 | +66:29:27.9 | 0.00 | 0.00 | 3.087 | 0.028 | 3.269 | 0.065 | | | |
| 282 | NEP180053+663942 | 18:00:53.3 | +66:39:42.5 | 0.01 | 0.01 | 0.194 | 0.029 | 0.103 | 0.029 | | | |
| 283 | NEP180053+662302 | 18:00:53.9 | +66:23:02.3 | 0.03 | 0.04 | 0.130 | 0.026 | 0.172 | 0.026 | 20.9 | | 31.1 |
| 284 | NEP180054+665443 | 18:00:54.6 | +66:54:43.7 | 0.08 | 0.18 | 0.197 | 0.040 | 0.252 | 0.042 | 20.1 | | 20.1 |
| 285 | NEP180054+662955 | 18:00:54.6 | +66:29:55.7 | 0.14 | 0.14 | 0.147 | 0.024 | 0.226 | 0.027 | | | |
| 286 | NEP180056+664225 | 18:00:56.1 | +66:42:25.3 | 0.01 | 0.03 | 0.172 | 0.026 | 0.148 | 0.026 | | | |
| 287 | NEP180057+663722 | 18:00:57.4 | +66:37:22.5 | 0.00 | 0.00 | 0.181 | 0.026 | 0.125 | 0.026 | | | |
| 288 | NEP180057+661200 | 18:00:57.8 | +66:12:00.1 | 0.02 | 0.02 | 0.373 | 0.036 | 0.308 | 0.037 | | | |
| 289 | NEP180059+661057 | 18:00:59.6 | +66:10:57.4 | 0.04 | 0.09 | 0.219 | 0.036 | 0.251 | 0.038 | | | |
| 290 | NEP180059+662546 | 18:00:59.8 | +66:25:46.2 | 0.07 | 0.12 | 0.124 | 0.027 | 0.195 | 0.028 | 21.1 | 7.7 | -24.1 |
| 291 | NEP180104+662738 | 18:01:04.3 | +66:27:38.3 | 0.00 | 0.00 | 0.628 | 0.028 | 0.714 | 0.028 | | | |
| 292 | NEP180104+663549 | 18:01:04.9 | +66:35:50.0 | 0.01 | 0.01 | 0.259 | 0.026 | 0.220 | 0.026 | | | |
| 293 | NEP180105+665812 | 18:01:05.1 | +66:58:12.2 | 0.00 | 0.00 | 0.396 | 0.053 | 0.268 | 0.053 | | | |
| 294 | NEP180105+664413 | 18:01:05.3 | +66:44:13.4 | 0.27 | 0.20 | 0.123 | 0.026 | 0.285 | 0.027 | 26.9 | 16.5 | -64.1 |
| 295 | NEP180106+662753 | 18:01:06.4 | +66:27:53.2 | 0.06 | 0.07 | 0.659 | 0.028 | 0.789 | 0.043 | | | |
| 296 | NEP180106+664331 | 18:01:06.6 | +66:43:31.5 | 0.01 | 0.06 | 0.276 | 0.102 | 0.103 | 0.102 | | | |
| 297 | NEP180107+665910 | 18:01:07.3 | +66:59:10.4 | 0.00 | 0.00 | 2.227 | 0.053 | 2.198 | 0.056 | | | |
| 298 | NEP180108+664500 | 18:01:08.0 | +66:45:00.4 | 0.22 | 0.15 | 0.754 | 0.050 | 3.346 | 0.088 | 36.0 | 29.7 | -71.5 |
| 299 | NEP180109+663331 | 18:01:09.1 | +66:33:31.1 | 0.03 | 0.06 | 0.504 | 0.025 | 0.684 | 0.035 | 18.7 | | 8.1 |
| 300 | NEP180111+664022 | 18:01:11.4 | +66:40:22.4 | 0.02 | 0.03 | 0.237 | 0.031 | 0.134 | 0.032 | | | |

Table 1. – continued

| Running number | Source name | RA hh:mm:ss.s | DEC dd:mm:ss.s | δRA " | δDEC " | $S_{peak}$ mJy beam$^{-1}$ | $S_{peak}$ error mJy beam$^{-1}$ | $S_{total}$ mJy | $S_{total}$ error mJy | $θ_{maj}$ " | $θ_{min}$ " | PA ° |
|---|---|---|---|---|---|---|---|---|---|---|---|---|
| (1) | (2) | (3) | (4) | (5) | (6) | (7) | (8) | (9) | (10) | (11) | (12) | (13) |
| 301 | NEP180114+663113 | 18:01:14.4 | +66:31:13.3 | 0.00 | 0.00 | 3.979 | 0.032 | 4.242 | 0.049 | | | |
| 302 | NEP180116+662404 | 18:01:16.3 | +66:24:04.4 | 0.07 | 0.24 | 0.171 | 0.027 | 0.176 | 0.030 | | | |
| 303 | NEP180119+663401 | 18:01:19.4 | +66:34:01.9 | 0.01 | 0.01 | 0.226 | 0.025 | 0.184 | 0.025 | | | |
| 304 | NEP180120+663722 | 18:01:20.5 | +66:37:22.1 | 0.01 | 0.02 | 0.207 | 0.025 | 0.144 | 0.025 | | | |
| 305 | NEP180121+663031 | 18:01:21.2 | +66:30:31.6 | 0.10 | 0.03 | 0.443 | 0.032 | 1.125 | 0.040 | 32.3 | 14.4 | 84.8 |
| 306 | NEP180121+671850 | 18:01:21.3 | +67:18:50.9 | 0.03 | 0.19 | 5.970 | 0.357 | 10.128 | 0.449 | 23.9 | 5.4 | 7.5 |
| 307 | NEP180123+664346 | 18:01:23.8 | +66:43:46.3 | 0.02 | 0.01 | 12.286 | 0.102 | 18.249 | 0.460 | | | |
| 308 | NEP180127+664020 | 18:01:27.4 | +66:40:20.7 | 0.00 | 0.01 | 0.511 | 0.031 | 0.363 | 0.032 | | | |
| 309 | NEP180127+670054 | 18:01:27.7 | +67:00:54.1 | 0.03 | 0.04 | 0.489 | 0.061 | 0.708 | 0.063 | 19.4 | 6.5 | -26.3 |
| 310 | NEP180128+662854 | 18:01:28.2 | +66:28:54.1 | 0.01 | 0.01 | 0.667 | 0.031 | 0.722 | 0.034 | | | |
| 311 | NEP180128+671109 | 18:01:28.9 | +67:11:09.1 | 0.07 | 0.12 | 0.939 | 0.194 | 1.322 | 0.202 | 18.7 | 5.7 | 9.5 |
| 312 | NEP180133+664449 | 18:01:33.2 | +66:44:49.7 | 0.06 | 0.04 | 1.480 | 0.050 | 6.430 | 0.095 | 37.3 | 27.7 | 85.8 |
| 313 | NEP180133+665640 | 18:01:33.5 | +66:56:41.0 | 0.05 | 0.09 | 0.736 | 0.054 | 0.951 | 0.069 | | | |
| 314 | NEP180142+660416 | 18:01:42.1 | +66:04:16.4 | 0.01 | 0.02 | 0.361 | 0.065 | 0.276 | 0.065 | | | |
| 315 | NEP180143+664108 | 18:01:43.0 | +66:41:08.5 | 0.00 | 0.00 | 1.559 | 0.028 | 1.488 | 0.030 | | | |
| 316 | NEP180143+665252 | 18:01:43.7 | +66:52:52.5 | 0.00 | 0.00 | 1.026 | 0.037 | 1.037 | 0.039 | | | |
| 317 | NEP180143+664521 | 18:01:44.0 | +66:45:21.2 | 0.01 | 0.02 | 0.149 | 0.028 | 0.151 | 0.028 | | | |
| 318 | NEP180144+664137 | 18:01:44.5 | +66:41:37.6 | 0.00 | 0.00 | 0.339 | 0.031 | 0.264 | 0.031 | | | |
| 319 | NEP180146+661609 | 18:01:46.1 | +66:16:010.0 | 0.01 | 0.01 | 0.241 | 0.033 | 0.158 | 0.033 | | | |
| 320 | NEP180146+663840 | 18:01:46.8 | +66:38:40.5 | 0.00 | 0.00 | 1.436 | 0.028 | 1.419 | 0.035 | | | |

**Table 1.** *– continued*

| Running number | Source name | RA hh:mm:ss.s | DEC dd:mm:ss.s | δRA " | δDEC " | $S_{peak}$ mJy beam$^{-1}$ | $S_{peak}$ error mJy beam$^{-1}$ | $S_{total}$ mJy | $S_{total}$ error mJy | $\theta_{maj}$ " | $\theta_{min}$ " | PA ° |
|---|---|---|---|---|---|---|---|---|---|---|---|---|
| (1) | (2) | (3) | (4) | (5) | (6) | (7) | (8) | (9) | (10) | (11) | (12) | (13) |
| 321 | NEP180148+661438 | 18:01:48.6 | +66:14:39.0 | 0.00 | 0.00 | 0.623 | 0.039 | 0.554 | 0.039 | | | |
| 322 | NEP180148+670146 | 18:01:49.0 | +67:01:46.6 | 0.01 | 0.03 | 0.866 | 0.057 | 1.045 | 0.064 | | | |
| 323 | NEP180149+665912 | 18:01:49.2 | +66:59:12.3 | 0.01 | 0.01 | 0.511 | 0.057 | 0.391 | 0.057 | | | |
| 324 | NEP180149+665130 | 18:01:49.7 | +66:51:30.6 | 0.02 | 0.03 | 0.262 | 0.037 | 0.253 | 0.038 | | | |
| 325 | NEP180151+663424 | 18:01:51.2 | +66:34:24.4 | 0.00 | 0.00 | 0.330 | 0.027 | 0.222 | 0.027 | | | |
| 326 | NEP180152+664557 | 18:01:52.4 | +66:45:57.3 | 0.00 | 0.01 | 0.554 | 0.028 | 0.487 | 0.029 | | | |
| 327 | NEP180153+670608 | 18:01:53.7 | +67:06:08.9 | 0.00 | 0.00 | 3.660 | 0.115 | 4.192 | 0.120 | | | |
| 328 | NEP180154+661859 | 18:01:54.6 | +66:18:59.5 | 0.01 | 0.01 | 0.163 | 0.029 | 0.126 | 0.029 | | | |
| 329 | NEP180157+665717 | 18:01:57.7 | +66:57:17.5 | 0.02 | 0.02 | 6.969 | 0.106 | 10.952 | 0.260 | 18.5 | 8.3 | 40.7 |
| 330 | NEP180158+670732 | 18:01:58.7 | +67:07:32.3 | 0.10 | 0.64 | 0.518 | 0.115 | 1.928 | 0.122 | 47.6 | 15.9 | -5.1 |
| 331 | NEP18020+662526 | 18:02:0.3 | +66:25:26.3 | 0.02 | 0.02 | 0.157 | 0.027 | 0.126 | 0.027 | | | |
| 332 | NEP18020+663351 | 18:02:0.6 | +66:33:51.4 | 0.06 | 0.05 | 0.139 | 0.027 | 0.139 | 0.028 | | | |
| 333 | NEP180201+661647 | 18:02:01.3 | +66:16:47.4 | 0.06 | 0.08 | 0.368 | 0.033 | 0.423 | 0.038 | | | |
| 334 | NEP180204+665755 | 18:02:04.4 | +66:57:55.9 | 0.00 | 0.00 | 11.014 | 0.106 | 15.851 | 0.205 | | | |
| 335 | NEP180206+662608 | 18:02:06.0 | +66:26:08.5 | 0.03 | 0.04 | 0.404 | 0.028 | 0.428 | 0.032 | | | |
| 336 | NEP180208+661748 | 18:02:08.5 | +66:17:48.1 | 0.01 | 0.01 | 0.515 | 0.037 | 0.469 | 0.039 | | | |
| 337 | NEP180216+664330 | 18:02:16.0 | +66:43:30.0 | 0.02 | 0.02 | 0.282 | 0.035 | 0.315 | 0.036 | | | |
| 338 | NEP180216+661903 | 18:02:17.0 | +66:19:03.9 | 0.02 | 0.04 | 0.170 | 0.036 | 0.302 | 0.036 | 23.0 | 11.1 | -16.7 |
| 339 | NEP180221+660032 | 18:02:21.4 | +66:00:32.4 | 0.00 | 0.01 | 4.310 | 0.097 | 5.088 | 0.128 | | | |
| 340 | NEP180223+671026 | 18:02:23.6 | +67:10:26.2 | 0.03 | 0.06 | 3.491 | 0.140 | 6.111 | 0.223 | 23.1 | 7.4 | 17.4 |

Table 1. – *continued*

| Running number | Source name | RA hh:mm:ss.s | DEC dd:mm:ss.s | $\delta$RA " | $\delta$DEC " | $S_{peak}$ mJy beam$^{-1}$ | $S_{peak}$ error mJy beam$^{-1}$ | $S_{total}$ mJy | $S_{total}$ error mJy | $\theta_{maj}$ " | $\theta_{min}$ " | PA ° |
|---|---|---|---|---|---|---|---|---|---|---|---|---|
| (1) | (2) | (3) | (4) | (5) | (6) | (7) | (8) | (9) | (10) | (11) | (12) | (13) |
| 341 | NEP180226+664105 | 18:02:26.9 | +66:41:05.8 | 0.01 | 0.01 | 0.462 | 0.031 | 0.426 | 0.032 | | | |
| 342 | NEP180229+664018 | 18:02:29.3 | +66:40:18.9 | 0.05 | 0.05 | 0.194 | 0.031 | 0.171 | 0.033 | | | |
| 343 | NEP180230+660937 | 18:02:30.1 | +66:09:37.8 | 0.06 | 0.04 | 0.337 | 0.049 | 0.252 | 0.051 | | | |
| 344 | NEP180232+665128 | 18:02:32.2 | +66:51:28.2 | 0.00 | 0.01 | 0.367 | 0.039 | 0.259 | 0.039 | | | |
| 345 | NEP180238+665536 | 18:02:38.7 | +66:55:36.4 | 0.00 | 0.00 | 0.455 | 0.047 | 0.426 | 0.049 | | | |
| 346 | NEP180242+662349 | 18:02:42.8 | +66:23:49.7 | 0.01 | 0.02 | 0.432 | 0.039 | 0.435 | 0.040 | | | |
| 347 | NEP180245+664158 | 18:02:45.6 | +66:41:58.8 | 0.00 | 0.00 | 0.520 | 0.035 | 0.420 | 0.036 | | | |
| 348 | NEP180246+655144 | 18:02:46.7 | +65:51:44.3 | 6.36 | 0.93 | 1.037 | 0.249 | 2.025 | 0.280 | 34.1 | 2.7 | 69.7 |
| 349 | NEP180246+664815 | 18:02:46.7 | +66:48:15.1 | 0.00 | 0.00 | 4.342 | 0.040 | 4.734 | 0.045 | | | |
| 350 | NEP180247+670015 | 18:02:47.9 | +67:00:15.2 | 0.18 | 0.36 | 0.280 | 0.064 | 0.576 | 0.069 | 27.1 | 11.0 | -19.0 |
| 351 | NEP180249+665106 | 18:02:49.4 | +66:51:06.8 | 0.00 | 0.01 | 0.273 | 0.040 | 0.246 | 0.040 | | | |
| 352 | NEP180252+670409 | 18:02:52.4 | +67:04:09.5 | 0.06 | 0.04 | 0.453 | 0.086 | 0.374 | 0.088 | | | |
| 353 | NEP180252+662055 | 18:02:52.7 | +66:20:55.8 | 0.02 | 0.05 | 0.229 | 0.041 | 0.165 | 0.042 | | | |
| 354 | NEP180253+665131 | 18:02:53.2 | +66:51:31.9 | 0.00 | 0.01 | 0.302 | 0.040 | 0.232 | 0.040 | | | |
| 355 | NEP180253+670331 | 18:02:53.6 | +67:03:31.9 | 0.40 | 0.15 | 0.398 | 0.086 | 0.582 | 0.091 | 21.4 | 5.5 | 67.1 |
| 356 | NEP180256+662232 | 18:02:56.9 | +66:22:32.4 | 0.05 | 0.05 | 0.211 | 0.039 | 0.145 | 0.040 | | | |
| 357 | NEP180258+662811 | 18:02:58.6 | +66:28:11.2 | 0.00 | 0.00 | 0.268 | 0.030 | 0.163 | 0.030 | | | |
| 358 | NEP18030+660412 | 18:03:0.3 | +66:04:12.4 | 0.03 | 0.02 | 0.532 | 0.084 | 0.437 | 0.085 | | | |
| 359 | NEP180301+662353 | 18:03:01.2 | +66:23:53.1 | 0.00 | 0.00 | 2.519 | 0.039 | 2.611 | 0.046 | | | |
| 360 | NEP180304+663634 | 18:03:04.7 | +66:36:35.0 | 0.07 | 0.09 | 0.190 | 0.031 | 0.210 | 0.033 | | | |

Table 1. – *continued*

| Running number | Source name | RA hh:mm:ss.s | DEC dd:mm:ss.s | δRA " | δDEC " | $S_{peak}$ mJy beam$^{-1}$ | $S_{peak}$ error mJy beam$^{-1}$ | $S_{total}$ mJy | $S_{total}$ error mJy | $\theta_{maj}$ " | $\theta_{min}$ " | PA ° |
|---|---|---|---|---|---|---|---|---|---|---|---|---|
| (1) | (2) | (3) | (4) | (5) | (6) | (7) | (8) | (9) | (10) | (11) | (12) | (13) |
| 361 | NEP180311+663848 | 18:03:11.3 | +66:38:48.8 | 0.00 | 0.00 | 1.122 | 0.046 | 1.049 | 0.048 | | | |
| 362 | NEP180311+661408 | 18:03:11.8 | +66:14:08.2 | 0.13 | 0.18 | 0.259 | 0.052 | 0.288 | 0.054 | | | |
| 363 | NEP180312+660147 | 18:03:12.3 | +66:01:48.0 | 0.03 | 0.04 | 0.871 | 0.128 | 1.274 | 0.132 | 17.7 | 9.5 | -35.5 |
| 364 | NEP180312+662030 | 18:03:12.8 | +66:20:30.4 | 0.05 | 0.06 | 0.234 | 0.041 | 0.162 | 0.042 | | | |
| 365 | NEP180314+655423 | 18:03:14.4 | +65:54:23.8 | 0.01 | 0.02 | 5.138 | 0.333 | 6.956 | 0.361 | 17.9 | 4.1 | -25.7 |
| 366 | NEP180316+664226 | 18:03:16.3 | +66:42:26.5 | 0.00 | 0.00 | 8.307 | 0.052 | 8.565 | 0.081 | | | |
| 367 | NEP180321+663052 | 18:03:21.4 | +66:30:52.2 | 0.01 | 0.01 | 0.247 | 0.041 | 0.169 | 0.041 | | | |
| 368 | NEP180328+664108 | 18:03:28.7 | +66:41:08.9 | 0.01 | 0.01 | 0.581 | 0.046 | 0.671 | 0.048 | | | |
| 369 | NEP180329+664707 | 18:03:29.2 | +66:47:07.3 | 0.19 | 0.25 | 0.231 | 0.038 | 0.405 | 0.041 | 21.4 | 11.5 | -41.4 |
| 370 | NEP180330+661512 | 18:03:30.1 | +66:15:13.0 | 0.04 | 0.05 | 0.246 | 0.045 | 0.196 | 0.046 | | | |
| 371 | NEP180331+662112 | 18:03:31.4 | +66:21:12.6 | 0.00 | 0.00 | 0.468 | 0.042 | 0.389 | 0.042 | | | |
| 372 | NEP180334+662033 | 18:03:34.8 | +66:20:33.9 | 0.02 | 0.01 | 0.300 | 0.051 | 0.270 | 0.051 | | | |
| 373 | NEP180341+670310 | 18:03:41.7 | +67:03:10.4 | 0.19 | 0.30 | 0.530 | 0.134 | 2.483 | 0.144 | 38.4 | 31.4 | 3.3 |
| 374 | NEP180346+662054 | 18:03:46.6 | +66:20:54.3 | 0.01 | 0.01 | 0.733 | 0.051 | 0.623 | 0.053 | | | |
| 375 | NEP180346+660927 | 18:03:46.9 | +66:09:27.8 | 0.01 | 0.01 | 0.837 | 0.078 | 0.777 | 0.079 | | | |
| 376 | NEP180347+663437 | 18:03:47.9 | +66:34:37.4 | 0.02 | 0.02 | 0.212 | 0.033 | 0.131 | 0.034 | | | |
| 377 | NEP180348+660540 | 18:03:49.0 | +66:05:40.8 | 0.01 | 0.02 | 9.358 | 0.154 | 13.784 | 0.205 | 18.8 | 6.4 | -33.3 |
| 378 | NEP180351+660516 | 18:03:51.2 | +66:05:16.8 | 0.00 | 0.01 | 27.202 | 0.154 | 41.616 | 0.611 | 20.4 | | -25.9 |
| 379 | NEP180352+655531 | 18:03:52.9 | +65:55:31.5 | 0.00 | 0.00 | 1.155 | 0.225 | 1.074 | 0.235 | | | |
| 380 | NEP180355+662034 | 18:03:55.5 | +66:20:34.8 | 0.16 | 0.21 | 0.217 | 0.051 | 0.470 | 0.054 | 28.3 | 12.5 | -36.8 |

Table 1. – continued

| Running number | Source name | RA hh:mm:ss.s | DEC dd:mm:ss.s | δRA " | δDEC " | $S_{peak}$ mJy beam$^{-1}$ | $S_{peak}$ error mJy beam$^{-1}$ | $S_{total}$ mJy | $S_{total}$ error mJy | $\theta_{maj}$ " | $\theta_{min}$ " | PA ° |
|---|---|---|---|---|---|---|---|---|---|---|---|---|
| (1) | (2) | (3) | (4) | (5) | (6) | (7) | (8) | (9) | (10) | (11) | (12) | (13) |
| 381 | NEP180357+662735 | 18:03:57.7 | +66:27:35.0 | 0.03 | 0.05 | 0.286 | 0.043 | 0.274 | 0.045 | | | |
| 382 | NEP18040+662836 | 18:04:0.3 | +66:28:36.8 | 0.06 | 0.03 | 0.279 | 0.043 | 0.264 | 0.045 | | | |
| 383 | NEP18040+663525 | 18:04:0.9 | +66:35:25.6 | 0.06 | 0.05 | 0.695 | 0.033 | 0.989 | 0.051 | | | |
| 384 | NEP180403+671318 | 18:04:03.8 | +67:13:18.3 | 0.05 | 0.07 | 1.553 | 0.283 | 1.356 | 0.292 | | | |
| 385 | NEP180407+671402 | 18:04:08.0 | +67:14:02.4 | 0.04 | 0.08 | 3.657 | 0.491 | 7.951 | 0.527 | 28.0 | 12.0 | -4.8 |
| 386 | NEP180408+664426 | 18:04:08.4 | +66:44:26.2 | 0.01 | 0.01 | 0.556 | 0.050 | 0.518 | 0.051 | | | |
| 387 | NEP180410+655149 | 18:04:10.4 | +65:51:49.6 | 0.11 | 0.27 | 2.039 | 0.468 | 2.605 | 0.489 | 19.1 | | 26.7 |
| 388 | NEP180410+662723 | 18:04:10.5 | +66:27:23.3 | 0.06 | 0.06 | 0.826 | 0.053 | 1.476 | 0.068 | 25.3 | 7.1 | 44.9 |
| 389 | NEP180410+665712 | 18:04:10.7 | +66:57:12.1 | 0.05 | 0.10 | 2.492 | 0.092 | 5.091 | 0.114 | 24.5 | 12.6 | 11.7 |
| 390 | NEP180411+665737 | 18:04:11.7 | +66:57:37.7 | 0.05 | 0.41 | 3.504 | 0.092 | 9.991 | 0.296 | 41.8 | 7.2 | 9.6 |
| 391 | NEP180422+660021 | 18:04:22.2 | +66:00:21.0 | 0.03 | 0.02 | 1.153 | 0.173 | 0.951 | 0.178 | | | |
| 392 | NEP180422+664132 | 18:04:23.0 | +66:41:32.2 | 0.04 | 0.04 | 0.267 | 0.047 | 0.210 | 0.048 | | | |
| 393 | NEP180425+664627 | 18:04:26.0 | +66:46:27.4 | 0.09 | 0.08 | 0.313 | 0.050 | 0.292 | 0.053 | | | |
| 394 | NEP180426+662244 | 18:04:26.5 | +66:22:44.3 | 0.04 | 0.04 | 0.320 | 0.053 | 0.383 | 0.055 | | | |
| 395 | NEP180427+662403 | 18:04:27.5 | +66:24:03.5 | 0.42 | 0.26 | 0.249 | 0.049 | 0.396 | 0.055 | 19.8 | 10.3 | -70.8 |
| 396 | NEP180435+662707 | 18:04:35.9 | +66:27:07.6 | 0.00 | 0.00 | 6.454 | 0.061 | 6.697 | 0.084 | | | |
| 397 | NEP180436+662531 | 18:04:36.8 | +66:25:32.0 | 0.00 | 0.00 | 73.613 | 0.144 | 77.644 | 0.394 | | | |
| 398 | NEP180438+660500 | 18:04:38.9 | +66:05:00.0 | 0.00 | 0.00 | 3.474 | 0.175 | 4.146 | 0.182 | | | |
| 399 | NEP180453+664021 | 18:04:53.5 | +66:40:21.9 | 0.03 | 0.02 | 0.276 | 0.052 | 0.281 | 0.054 | | | |
| 400 | NEP180455+665251 | 18:04:55.2 | +66:52:51.9 | 8.26 | 8.40 | 0.719 | 0.079 | 0.275 | 0.079 | | | |

Table 1. – *continued*

| Running number | Source name | RA hh:mm:ss.s | DEC dd:mm:ss.s | δRA " | δDEC " | $S_{peak}$ mJy beam$^{-1}$ | $S_{peak}$ error mJy beam$^{-1}$ | $S_{total}$ mJy | $S_{total}$ error mJy | $\theta_{maj}$ " | $\theta_{min}$ " | PA ° |
|---|---|---|---|---|---|---|---|---|---|---|---|---|
| (1) | (2) | (3) | (4) | (5) | (6) | (7) | (8) | (9) | (10) | (11) | (12) | (13) |
| 401 | NEP180456+664633 | 18:04:56.2 | +66:46:33.5 | 0.14 | 0.14 | 0.252 | 0.057 | 0.357 | 0.059 | | | |
| 402 | NEP180456+663106 | 18:04:56.3 | +66:31:06.5 | 0.09 | 0.07 | 0.310 | 0.060 | 0.462 | 0.061 | 19.3 | 8.7 | -51.9 |
| 403 | NEP180457+665351 | 18:04:57.8 | +66:53:51.5 | 0.02 | 0.02 | 16.561 | 0.787 | 34.084 | 0.972 | 21.0 | 16.5 | 26.1 |
| 404 | NEP180458+665700 | 18:04:58.1 | +66:57:00.8 | 0.01 | 0.02 | 8.422 | 0.109 | 18.625 | 0.328 | 23.6 | 14.7 | -0.1 |
| 405 | NEP180501+663050 | 18:05:01.2 | +66:30:50.7 | 0.11 | 0.17 | 0.780 | 0.060 | 0.985 | 0.086 | | | |
| 406 | NEP180502+665318 | 18:05:02.8 | +66:53:18.6 | 0.00 | 0.00 | 28.016 | 0.787 | 39.701 | 0.909 | | | |
| 407 | NEP180507+663132 | 18:05:07.8 | +66:31:32.7 | 0.08 | 0.15 | 0.259 | 0.054 | 0.396 | 0.056 | 19.9 | 7.8 | 3.6 |
| 408 | NEP180508+663239 | 18:05:08.3 | +66:32:39.2 | 0.25 | 0.21 | 0.323 | 0.066 | 0.548 | 0.072 | 23.3 | 8.7 | -54.2 |
| 409 | NEP180509+664533 | 18:05:09.2 | +66:45:33.6 | 0.17 | 0.25 | 0.313 | 0.073 | 0.738 | 0.076 | 31.3 | 12.9 | -33.6 |
| 410 | NEP180512+661957 | 18:05:12.5 | +66:19:57.7 | 0.01 | 0.01 | 0.607 | 0.095 | 0.472 | 0.096 | | | |
| 411 | NEP180514+665643 | 18:05:14.6 | +66:56:43.7 | 0.01 | 0.01 | 1.463 | 0.137 | 1.459 | 0.140 | | | |
| 412 | NEP180516+662111 | 18:05:16.0 | +66:21:11.6 | 0.01 | 0.01 | 5.583 | 0.078 | 9.464 | 0.166 | 19.9 | 9.6 | 16.2 |
| 413 | NEP180517+665403 | 18:05:17.2 | +66:54:03.6 | 0.01 | 0.01 | 1.475 | 0.109 | 1.554 | 0.114 | | | |
| 414 | NEP180517+664451 | 18:05:17.7 | +66:44:51.0 | 0.01 | 0.04 | 2.189 | 0.073 | 3.707 | 0.114 | 22.9 | 5.9 | -2.4 |
| 415 | NEP180518+663512 | 18:05:18.2 | +66:35:12.4 | 0.00 | 0.00 | 1.423 | 0.066 | 1.500 | 0.070 | | | |
| 416 | NEP180526+662105 | 18:05:26.6 | +66:21:05.2 | 0.08 | 0.06 | 0.444 | 0.078 | 0.664 | 0.082 | | | |
| 417 | NEP180527+663946 | 18:05:27.8 | +66:39:46.6 | 0.05 | 0.03 | 0.362 | 0.062 | 0.310 | 0.063 | | | |
| 418 | NEP180530+664451 | 18:05:30.9 | +66:44:51.5 | 0.01 | 0.01 | 1.928 | 0.073 | 2.299 | 0.084 | | | |
| 419 | NEP180531+671009 | 18:05:31.1 | +67:10:09.4 | 0.01 | 0.01 | 16.566 | 0.436 | 36.497 | 0.598 | 24.0 | 15.8 | 34.3 |
| 420 | NEP180532+661803 | 18:05:32.8 | +66:18:03.3 | 0.03 | 0.02 | 2.348 | 0.103 | 3.420 | 0.138 | 19.9 | 5.2 | -58.4 |

Table 1. −continued

| Running number | Source name | RA hh:mm:ss.s | DEC dd:mm:ss.s | δRA " | δDEC " | $S_{peak}$ mJy beam$^{-1}$ | $S_{peak}$ error mJy beam$^{-1}$ | $S_{total}$ mJy | $S_{total}$ error mJy | $\theta_{maj}$ " | $\theta_{min}$ " | PA ° |
|---|---|---|---|---|---|---|---|---|---|---|---|---|
| (1) | (2) | (3) | (4) | (5) | (6) | (7) | (8) | (9) | (10) | (11) | (12) | (13) |
| 421 | NEP180538+664157 | 18:05:38.5 | +66:41:57.4 | 0.00 | 0.00 | 1.383 | 0.083 | 1.327 | 0.084 | | | |
| 422 | NEP180541+664239 | 18:05:41.4 | +66:42:39.8 | 0.09 | 0.04 | 0.401 | 0.083 | 0.502 | 0.085 | | | |
| 423 | NEP180541+665002 | 18:05:41.7 | +66:50:02.8 | 0.01 | 0.02 | 0.932 | 0.121 | 0.972 | 0.122 | | | |
| 424 | NEP180545+662125 | 18:05:45.2 | +66:21:25.4 | 0.05 | 0.05 | 0.477 | 0.078 | 0.413 | 0.081 | | | |
| 425 | NEP180545+660331 | 18:05:45.4 | +66:03:31.0 | 0.15 | 0.07 | 1.288 | 0.255 | 1.033 | 0.263 | | | |
| 426 | NEP180545+662611 | 18:05:45.4 | +66:26:11.6 | 0.00 | 0.00 | 4.383 | 0.079 | 4.748 | 0.089 | | | |
| 427 | NEP180554+664525 | 18:05:54.1 | +66:45:25.1 | 0.01 | 0.01 | 6.323 | 0.093 | 8.993 | 0.156 | | | |
| 428 | NEP180556+663056 | 18:05:56.3 | +66:30:56.2 | 0.00 | 0.00 | 8.669 | 0.075 | 10.389 | 0.128 | | | |
| 429 | NEP180557+662954 | 18:05:57.7 | +66:29:54.8 | 0.00 | 0.00 | 3.768 | 0.075 | 4.436 | 0.091 | | | |
| 430 | NEP180557+664509 | 18:05:57.8 | +66:45:09.5 | 0.01 | 0.01 | 9.320 | 0.093 | 12.653 | 0.262 | | | |
| 431 | NEP180558+665503 | 18:05:58.5 | +66:55:03.4 | 0.01 | 0.01 | 3.902 | 0.123 | 5.289 | 0.147 | | | |
| 432 | NEP180618+662754 | 18:06:18.0 | +66:27:54.8 | 0.03 | 0.16 | 0.457 | 0.083 | 0.447 | 0.086 | | | |
| 433 | NEP180619+664024 | 18:06:19.6 | +66:40:24.2 | 0.35 | 0.22 | 0.360 | 0.080 | 0.618 | 0.084 | 25.9 | 6.9 | -60.5 |
| 434 | NEP180620+663006 | 18:06:20.9 | +66:30:06.2 | 0.30 | 4.99 | 1.001 | 0.090 | 2.792 | 0.175 | 44.4 | 3.7 | -2.5 |
| 435 | NEP180621+665121 | 18:06:21.5 | +66:51:21.1 | 0.13 | 0.10 | 0.641 | 0.135 | 0.709 | 0.140 | | | |
| 436 | NEP180625+661958 | 18:06:25.6 | +66:19:58.2 | 0.02 | 0.01 | 1.417 | 0.141 | 1.921 | 0.145 | | | |
| 437 | NEP180637+662203 | 18:06:37.9 | +66:22:03.1 | 0.10 | 0.88 | 0.598 | 0.145 | 2.044 | 0.157 | 53.5 | 8.5 | -1.5 |
| 438 | NEP180639+663243 | 18:06:39.5 | +66:32:43.1 | 0.00 | 0.00 | 1.268 | 0.130 | 1.394 | 0.131 | | | |
| 439 | NEP180642+662654 | 18:06:42.5 | +66:26:54.1 | 0.11 | 6.02 | 0.501 | 0.130 | 2.253 | 0.144 | 91.3 | | 2.9 |
| 440 | NEP180646+662738 | 18:06:46.9 | +66:27:38.2 | 0.00 | 0.00 | 7.998 | 0.130 | 8.760 | 0.147 | | | |

Table 1. –*continued*

| Running number | Source name | RA hh:mm:ss.s | DEC dd:mm:ss.s | δRA " | δDEC " | $S_{peak}$ mJy beam$^{-1}$ | $S_{peak}$ error mJy beam$^{-1}$ | $S_{total}$ mJy | $S_{total}$ error mJy | $\theta_{maj}$ " | $\theta_{min}$ " | PA ° |
|---|---|---|---|---|---|---|---|---|---|---|---|---|
| (1) | (2) | (3) | (4) | (5) | (6) | (7) | (8) | (9) | (10) | (11) | (12) | (13) |
| 441 | NEP180648+665041 | 18:06:48.5 | +66:50:41.8 | 0.03 | 0.06 | 1.178 | 0.214 | 1.257 | 0.219 | | | |
| 442 | NEP18070+665648 | 18:07:0.6 | +66:56:48.5 | 0.51 | 0.08 | 1.155 | 0.256 | 1.956 | 0.270 | 27.3 | 5.6 | 89.6 |
| 443 | NEP180701+664700 | 18:07:01.0 | +66:47:00.4 | 0.01 | 0.01 | 1.738 | 0.138 | 1.831 | 0.142 | | | |
| 444 | NEP180703+670115 | 18:07:03.3 | +67:01:15.6 | 0.12 | 0.17 | 1.901 | 0.378 | 2.904 | 0.398 | 18.1 | 10.0 | -8.6 |
| 445 | NEP180704+662807 | 18:07:04.5 | +66:28:07.8 | 0.03 | 0.03 | 6.598 | 0.160 | 10.341 | 0.343 | | | |
| 446 | NEP180714+660732 | 18:07:14.9 | +66:07:32.0 | 0.04 | 0.11 | 3.046 | 0.513 | 1.945 | 0.536 | | | |
| 447 | NEP180719+661847 | 18:07:19.0 | +66:18:47.1 | 0.11 | 0.34 | 6.731 | 0.232 | 17.190 | 0.791 | 30.4 | 13.0 | -5.5 |
| 448 | NEP180720+661452 | 18:07:20.2 | +66:14:52.9 | 0.19 | 0.29 | 1.132 | 0.239 | 1.627 | 0.254 | 20.6 | 3.8 | -34.1 |
| 449 | NEP180722+662306 | 18:07:22.9 | +66:23:06.3 | 0.04 | 0.03 | 1.483 | 0.180 | 2.155 | 0.186 | | | |
| 450 | NEP180729+662836 | 18:07:29.6 | +66:28:36.6 | 0.03 | 0.05 | 3.280 | 0.201 | 8.399 | 0.245 | 32.6 | 13.5 | 32.1 |
| 451 | NEP180736+665755 | 18:07:36.1 | +66:57:55.0 | 0.25 | 0.16 | 2.080 | 0.453 | 3.095 | 0.470 | 19.8 | 8.1 | 56.3 |
| 452 | NEP180741+663725 | 18:07:41.7 | +66:37:25.8 | 0.03 | 0.02 | 1.696 | 0.205 | 2.115 | 0.209 | | | |
| 453 | NEP180750+663130 | 18:07:50.7 | +66:31:30.6 | 0.04 | 0.08 | 1.311 | 0.242 | 1.072 | 0.248 | | | |
| 454 | NEP180751+662400 | 18:07:51.8 | +66:24:00.2 | 0.15 | 0.16 | 1.181 | 0.217 | 1.370 | 0.235 | | | |
| 455 | NEP180752+661737 | 18:07:52.9 | +66:17:37.6 | 0.01 | 0.01 | 12.760 | 0.325 | 18.159 | 0.454 | | | |
| 456 | NEP180757+662852 | 18:07:58.0 | +66:28:53.0 | 0.24 | 1.78 | 0.896 | 0.201 | 1.933 | 0.219 | 44.1 | | -10.7 |
| 457 | NEP180823+663709 | 18:08:24.0 | +66:37:09.4 | 0.10 | 0.42 | 1.153 | 0.272 | 2.601 | 0.284 | 35.4 | 6.7 | 17.7 |
| 458 | NEP180829+663840 | 18:08:30.0 | +66:38:40.3 | 0.05 | 0.13 | 1.723 | 0.298 | 1.280 | 0.313 | | | |
| 459 | NEP180849+663434 | 18:08:49.1 | +66:34:34.8 | 0.00 | 0.00 | 13.220 | 0.436 | 18.334 | 0.483 | | | |
| 460 | NEP180855+664617 | 18:08:55.4 | +66:46:17.7 | 0.51 | 0.92 | 2.688 | 0.522 | 4.104 | 0.554 | 21.1 | 5.3 | 0.8 |

Table 1. *—continued*

| Running number | Source name | RA hh:mm:ss.s | DEC dd:mm:ss.s | $\delta$RA " | $\delta$DEC " | $S_{peak}$ mJy beam$^{-1}$ | $S_{peak}$ error mJy beam$^{-1}$ | $S_{total}$ mJy | $S_{total}$ error mJy | $\theta_{maj}$ " | $\theta_{min}$ " | PA ° |
|---|---|---|---|---|---|---|---|---|---|---|---|---|
| (1) | (2) | (3) | (4) | (5) | (6) | (7) | (8) | (9) | (10) | (11) | (12) | (13) |
| 461 | NEP180859+663059 | 18:08:59.4 | +66:30:59.5 | 0.04 | 0.01 | 4.409 | 0.498 | 5.925 | 0.512 | 20.3 | 3.9 | -86.2 |
| 462 | NEP180910+663849 | 18:09:11.0 | +66:38:49.5 | 0.44 | 0.51 | 2.607 | 0.602 | 4.774 | 0.666 | 19.9 | 14.6 | -61.4 |